\documentclass[aps,pra,showpacs,twoside,twocolumn,10pt,nofootinbib]{revtex4-2}
\usepackage[utf8]{inputenc}
\usepackage[sc,osf]{mathpazo}
\usepackage{amsmath}
\usepackage[T1]{fontenc}
\usepackage{latexsym}
\usepackage{amssymb}
\usepackage{amsthm}
\usepackage[colorlinks=true,citecolor=blue,urlcolor=blue]{hyperref}
\usepackage{color}
\usepackage{graphics,epstopdf}
\usepackage{soul}
\usepackage[demo]{graphicx}
\usepackage{capt-of}
\usepackage{lipsum}
\usepackage{adjustbox}
\usepackage[normalem]{ulem}
\usepackage[table,xcdraw]{xcolor}
\usepackage[dvipsnames]{xcolor}
\usepackage{braket}
\usepackage{multirow}
\usepackage{physics}
\usepackage{ragged2e}
\usepackage{mathtools}
\usepackage{booktabs}

\usepackage{comment}
\usepackage[utf8]{inputenc}

\usepackage[font=small,labelfont=bf,justification=justified,singlelinecheck=false]{caption}

\definecolor{checkcolor}{HTML}{7541C0}

\definecolor{boldcolor}{HTML}{7541C0}
\definecolor{boldcolor1}{HTML}{4A2166}
\definecolor{boldcolor2}{HTML}{6C3391}
\definecolor{boldcolor3}{HTML}{366E8A}
\definecolor{boldcolor4}{HTML}{69B2D6}

\usepackage{calligra}
\DeclareMathAlphabet{\mathcalligra}{T1}{calligra}{m}{n}
\DeclareFontShape{T1}{calligra}{m}{n}{<->s*[2.2]callig15}{}

\newcommand{\THEO}[1]{\vspace{0.1cm}\noindent\textbf{\textcolor{boldcolor1}{$\blacksquare$ Theorem #1.}}}
\newcommand{\PROP}[1]{\vspace{0.1cm}\noindent\textbf{\textcolor{boldcolor2}{$\blacksquare$ Proposition #1.}}}
\newcommand{\CORL}[1]{\vspace{0.1cm}\noindent\textbf{\textcolor{boldcolor3}{$\Diamond$ Corollary #1.}}}

\begin{document}

\title{Dimensional advantage in network cooling with hybrid oscillator-qudit systems}

\author{Mrinmoy Samanta$^1$, Debkanta Ghosh$^1$, Rivu Gupta$^2$, Aditi Sen(De)$^1$}

\affiliation{$^1$ Harish-Chandra Research Institute, a CI of HBNI, Chhatnag Road, Jhunsi, Allahabad 211 019, India}

\affiliation{$^2$ Dipartimento di Fisica “Aldo Pontremoli,” Università degli Studi di Milano, I-$20133$ Milano, Italy}

\begin{abstract}

We examine the cooling of networks of oscillators through repeated unitary evolution followed by conditional measurement on a finite-dimensional auxiliary system, coupled via Jaynes-Cummings type interaction. We prove that near-perfect cooling of the oscillator to vacuum is fundamentally impossible when the auxiliary system is a qubit, establishing a no-cooling theorem for a two-level regulator. Moving beyond this limitation, we reveal a twofold {\it dimensional advantage} of higher-dimensional auxiliaries - reducing the number of required cycles, and enabling the efficient cooling of oscillators with higher initial energies. 
We further show that, while extending the network leads to a saturation of this dimensional advantage at moderate auxiliary dimensions, near-perfect cooling remains achievable for linear network configurations but fails for star networks. Moreover, we highlight the adaptability of the proposed protocol by demonstrating efficient cooling of hybrid continuous- and discrete-variable systems that naturally support the generation of non-Gaussian and entangled quantum resources.




\end{abstract}
\maketitle

\section{Introduction}
\label{sec:intro}


The miniaturization of thermal devices has enabled the realization of microscopic machines~\cite{Alicki_PRE_2013_battery, Joulain_PRL_2016_transistor, Ordonex-Miranda_PRE_2017_diode}, operating in the quantum regime, giving rise to the field of quantum thermodynamics~\cite{Gemmer_2004_quantum-thermodynamics, Binder_2018_quantum-thermodynamics, Deffner_2019_quantum-thermodynamics}. A central achievement in this direction is the quantum refrigerator~\cite{Allahverdyan_PRL_2004_refrigerator, Linden_PRL_2010_refrigerator, Elias_PRA_2011_refrigerator_algorithmic-cooling, Kosloff_ARPC_2014_refrigerator_review, Arsoy_SR_2021_few-qubit_refrigerator, Cangemi_PR_2024_refrigerator_review}, which can cool systems close to their ground state with efficiencies surpassing classical limits~\cite{Geva_JCP_1992_classical_thermodynamics_limits, Allahverdyan_PRE_2011_limits-of_cooling, Clivaz_PRL_2019_bounds_cooling}. Preparing states with low-entropy and high-purity is a prerequisite for numerous quantum information–theoretic tasks, including measurement-based quantum computation~\cite{Briegel_NP_2009_measurement-based_quantum-computation, Gu_PRA_2009_CV_cluster_quantum-computing}, boson sampling~\cite{Aaronson_PACM_2011_boson-sampling}, verifying the quantum advantage in computation~\cite{Pashayan_PRL_2015_simulation-complexity}, and implementing ultra-high-precision measurements~\cite{Caves_RMP_1980_ultrafast-measurement, Bocko_RMP_1996_ultrafast_measurement_experiment}. 

In discrete-variable (DV) platforms, quantum refrigeration has been widely investigated using two complementary approaches. The first relies on interacting spin systems~\cite{Sutherland1975, Takhtajan_PLA_1982_isotropic_Heisenberg-chain, Babujian_PLA_1982_exact_Heisenberg-chain, Fath_PRB_1991_bilinear-biquadratic_trapping, Fath_PRB_1993_isotropic_spin1}, 
ranging from qubits~\cite{Henrich_PRE_2007_driven-spin_refrigerator, Das_EPL_2019_transient_refrigerator, Hewgill_PRE_2020_three-qubit_refrigerator, Bhandari_PRB_2021_two-body_absorption-refrigerator, Konar_PRA_2022_robust_refrigerator_disorder} to higher-dimensional particles~\cite{Naseem_QST_2020_absorption-refrigerator_oscillator, Konar_PRA_2023_qudit_refrigerator, Ghosh_arXiv_2024_measurement_qudit_refrigerator}, using paradigms such as  self-contained engines~\cite{Linden_PRL_2010_refrigerator, Correa_PRE_2013_absorption-refrigerator_bound}, periodic driving~\cite{Niedenzu_NC_2018_refrigeration_periodic-driving}, repeated collisions~\cite{DeChiara_NJP_2018_refrigerator_repeated-collision}, and reverse-coupling mechanisms~\cite{Silva_PRE_2015_refrigerator_reverse-coupling}. The second approach is measurement-based refrigeration inspired by Zeno-type repeated measurements~\cite{Misra_JMP_1977_Zeno-paradox, Degasperis_INCA_1974_Zeno},  
where selective measurements on an auxiliary system indirectly purify the target by coupling it to an auxiliary system~\cite{Nakazato_PRL_2003_purification_repeated-measurement, Nakazato_PRA_2004_entanglement-purification_repeated-measurement, Nakazato_PRA_2008_repeated-measurement_decoherence, Bellomo_PRA_2010_distillation_repeated-measurement}. Although inherently probabilistic, this latter strategy has been shown to achieve efficient cooling in short- and variable-range interacting spin systems~\cite{Burgarth_PRA_2007_homogenization_repeated-measurement, Burgarth_arXiv_2007_cooling_controlling_local-interaction, Zhang_PRA_2019_cooling_repeated-measurement, Langbehn_arXiv_2005_cooling_random-measurement, Konar_PRA_2022_robust_refrigerator_disorder, Konar_PRA_2022_refrigeration_purification}, as well as chains of higher-dimensional quantum systems (qudits) through the subspace cooling technique~\cite{Ghosh_arXiv_2024_measurement_qudit_refrigerator}. Beyond refrigeration, Zeno-type protocols also find applications in generating both bipartite and multipartite entanglement~\cite{Wu_PRA_2004_long-range_entanglement_measurement, Qiu_PRA_2012_preparation_purification_entanglement_measurement, Yan_PRA_2023_eigenstate_preparation_measurement, Bera_arXiv_2005_volume-law-entanglement_random-measurement} and in designing quantum batteries~\cite{Yan_PRAppl_2023_charging_measurement, Chaki_PRA_2025_energy-distillation_measurement}, underscoring their versatility for quantum technologies.

On the other hand, for continuous-variable (CV) systems, cooling to the vacuum state is particularly crucial, as it enables universal quantum computation with Gaussian operations supplemented by Gottesman–Kitaev–Preskill states~\cite{Baragiola_PRL_2019_all-Gaussian_GKP}, increases the simulation complexity of practical CV  circuits~\cite{Calcluth_PRA_2023_vacuum_advantage_simulability}, and facilitates the preparation of nonclassical resources such as higher Fock states in superconducting circuits~\cite{Eickbusch_NP_2022_SNAP-gate, Kundra_PRX_2022_SNAP-displacement_state-preparation}.
Recent advances show that measurement-based protocols employing auxiliary DV regulators enable the preparation of Fock-state superpositions and efficient cooling~\cite{Zhang_PRA_2024_Fock-state_measurement, Zhang_PRA_2026_state-preparation_measurement}.  Efficient cooling of multiple CV systems, each coupled to specific transitions in an auxiliary qudit system, emerges as a direct consequence of the aforementioned scheme~\cite{Yan_PRA_2022_cooling_measurement}. Such systems leverage both discrete and continuous degrees of freedom and naturally operate in the hybrid CV–DV regime, which is ubiquitous in superconducting platforms~\cite{Andersen_NP_2015_hybrid-system, Hirayama_SNS_2021_hybrid-book, Kemper_arXiv_2025_hybrid-computing} and allows for scalable measurement-based cooling~\cite{Yan_PRA_2022_cooling_measurement} with more efficiency than other existing techniques~\cite{Genes_NJP_2008_sideband-cooling, Agarwal_PRA_2013_phonon-cooling, Ockeloen-Korppi_PRA_2019_sideband-cooling, Ockeloen-Korppi_PRA_2019_sideband-cooling, Qiao_PRL_2021_EIT_cooling}.  More broadly, hybrid systems provide a powerful platform for entanglement generation ~\cite{Kreis_PRA_2012_hybrid_entanglement, Jeong_NP_2014_hybrid_entanglement-generation} and distillation~\cite{Brask_PRL_2010_hybrid_entanglement-distribution}, quantum teleportation~\cite{Takeda_Nature_2013_hybrid_deterministic-teleportation, Andersen_PRL_2013_hybrid_CV-teleportation}, state-tomography~\cite{LinPeng_NJP_2013_hybrid_tomography, Wang_Science_2016_Schroedinger-cat_two-boxes}, quantum algorithms~\cite{Kang_arXiv_2023_hybrid_Trotter, Chakraborty_Quantum_2024_hybrid_linear-combination_unitary, Singh_arXiv_2025_hybrid_signal-processing}, quantum sensing~\cite{Rani_PRAppl_2025_hybrid_sensing_superconducting, Novikov_Nature_2025_hybrid_network-sensing, Naikoo_NJP_2025_hybrid_exceptional-sensing} and studying non-locality~\cite{Halder_PRA_2025_hybrid_nonlocality}.

In this work, we investigate the cooling of a network of continuous-variable quantum systems using repeated unitary evolution and conditional measurements performed on a coupled auxiliary {\it qudit}. Distinct from a related work~\cite{Yan_PRA_2022_cooling_measurement}, our model does not restrict the coupling of the CV oscillators to particular levels of the auxiliary qudit; instead, each oscillator interacts with the entire qudit through a Jaynes-Cummings (JC) type interaction. Our principal question of interest is, ``{\it can higher-dimensional discrete quantum systems provide greater advantage in cooling oscillator chains?}'' We report that the answer is affirmative, thereby extending the notion of {\it dimensional advantage}, previously observed in quantum key distribution~\cite{Bourennane_PRA_2001_high-d-QKD, Cerf_PRL_2002_high-d-BB84-six-state, Ogrodnik_arXiv_2024_high-d-QKD-resource-efficient, Patra_PRA_2024_qudit_SDC, Patra_arXiv_2025_qudit_LM05}, quantum batteries~\cite{Santos_PRE_2019_qudit_battery, Dou_EPL_2020_qutrit_battery, Ghosh_PRA_2022_qudit_battery_imperfection}, quantum cloning~\cite{Haw_NC_2016_hybrid_cloning}, and blind quantum computing~\cite{Romanova_arXiv_2025_qudit_blind-computing} to the domain of quantum refrigeration. Importantly, qudit-based protocols are experimentally accessible across diverse platforms, including twisted photons~\cite{Bouchard_Quantum_2018_twisted-photons}, telecommunication fibres~\cite{Canas_PRA_2017_telecommunication-fibre}, time-bin encoding~\cite{Islam_SA_2017_time-bin}, microwaves~\cite{Sit_OL_2018_microwave_qudit, Amitonova_OE_2020_microwave_qudit}, and silicon integrated-circuits~\cite{Ding_NPJ_2017_silicon-integrated}.

The fundamental origin of dimensional advantage that we report is a {\bf no-cooling theorem for qubits}: {\it it is not possible to cool even a single CV oscillator with near-perfect fidelity and non-vanishing success probability when the auxiliary system is a qubit.} This opens the door to examining the performance of DV auxiliaries of three levels (qutrits) and beyond. We quantify the cooling performance using the Uhlmann fidelity~\cite{Uhlmann_RMP_1976_fidelity, Jozsa_JMO_1994_mixed-state_fidelity, Mendonca_PRA_2008_fidelity, Miszczak_QIC_2009_fidelity-bounds}, initializing the oscillator in the most general single-mode Gaussian state, a displaced squeezed thermal state.
Our results emphasize a twofold dimensional advantage - firstly, the number of cycles of evolution and measurement, required for efficient cooling, decreases steadily with increasing auxiliary dimension; secondly, with increasing  dimension, it enables efficient cooling of oscillators with higher initial energies, which are otherwise difficult to cool due to their suppressed vacuum populations.
We first highlight the dimensional advantage concretely for a single oscillator and gain insights into the particulars of the cooling dynamics. We derive the optimal evolution time for auxiliary states of different dimensions and establish that the optimal time is determined solely by the desired measurement outcome, independent of the system state. Although the number of cycles required to achieve unit fidelity saturates as the auxiliary dimension increases, we prove that employing another oscillator as the auxiliary  can reduce this requirement to a single cycle. We then extend our protocol to cool multiple oscillators arranged in two distinct network configurations - the linear and the star networks. Interestingly, we show that while the linear network can admit near-perfect cooling with unit fidelity and moderate success probability,
star networks perform poorly even for small system sizes.
In the case of network cooling, increasing the auxiliary dimension indefinitely, i.e., using another oscillator as the auxiliary system, does not contribute much in terms of dimensional advantage, since the dynamics equilibrate to initial-state-dependent results for moderate qudit dimensions, up to eight. 

As the final piece in our puzzle, we analyze the cooling of a hybrid CV-DV system using a second qudit as the auxiliary. We demonstrate near-unit cooling fidelity for both subsystems when the auxiliary dimension is four or higher. Moreover, the dimensional advantage is further enhanced by higher-dimensional target qudits, which require fewer cooling cycles.
Further, we delineate applications of the final cooled state in generating non-Gaussian states essential for error-correction, such as CAT states, and also in producing hybrid CV-DV entanglement and $N00N$ states, with increasing dimension allowing access to higher excitation manifolds.

The remaining article is structured as follows. In Sec.~\ref{sec:cooling}, we describe the theoretical framework for cooling, and show that the dynamics may be captured through an effective evolution operator acting only on the system. We also derive conditions on the maximal achievable fidelity and its corresponding success probability in terms of the initial state of the system. Sec.~\ref{sec:single_oscilator_cooling} highlights the dimensional advantage of using auxiliary states of higher-than-qubit dimensions, while in Sec.~\ref{sec:multi_oscillator_cooling} we discuss the case for linear and star networks. Cooling of a hybrid CV-DV system is analyzed in Sec.~\ref{sec:hybrid-cooling} with possible applications of the final state being motivated. We end our study with conclusions in Sec.~\ref{sec:conclu}.

\section{Theoretical background of hybrid refrigeration}
\label{sec:cooling}

\begin{figure*}
\includegraphics [width=0.8\linewidth]{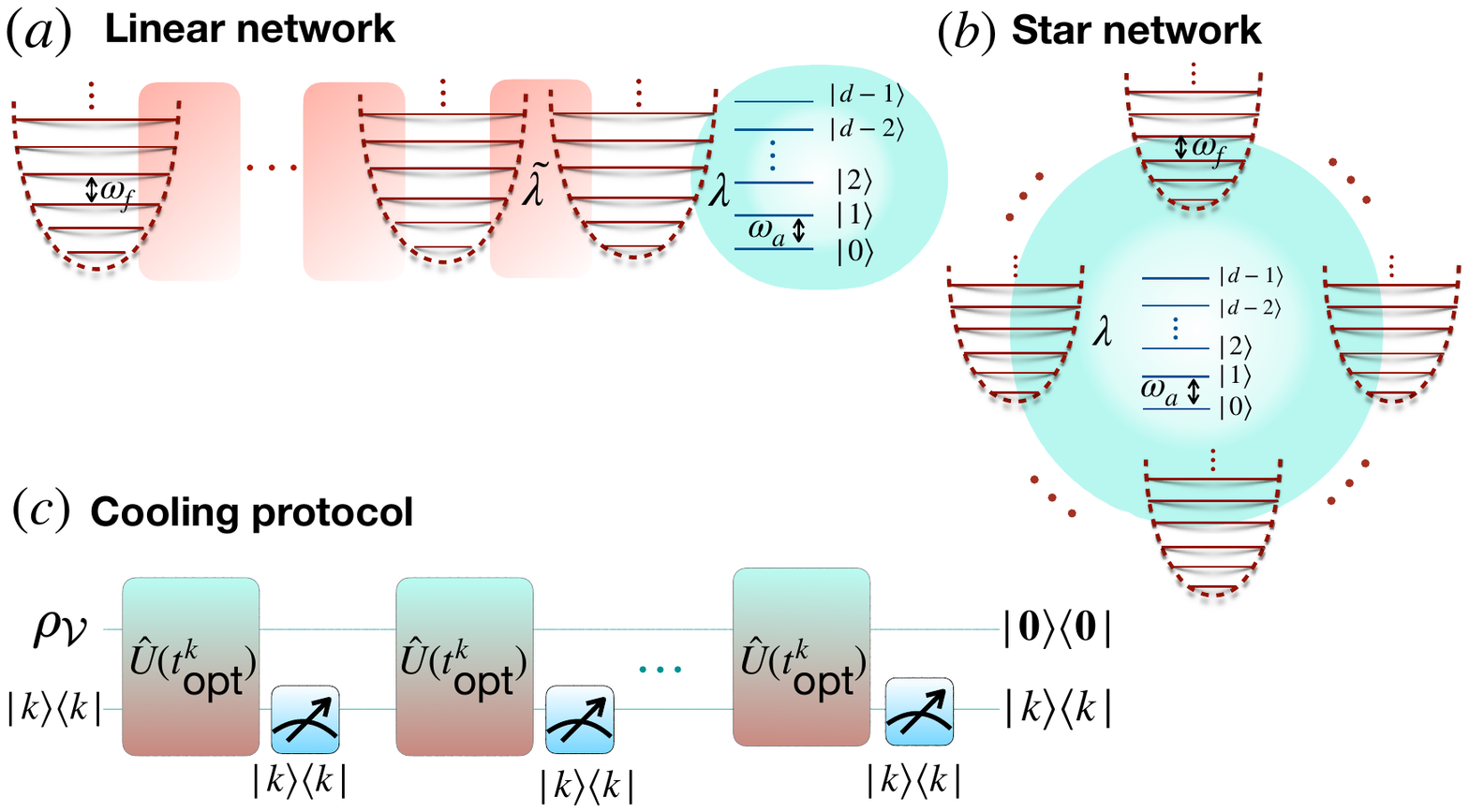 }
\caption{\justifying {\bf Schematic diagram illustrating different components of the cooling protocol.} $(a).$ Network of resonators, each with level spacing $\omega_f$, arranged in a linear configuration with the first resonator on the right coupled with strength $\lambda$ to the $d$-level regulator, with level spacing $\omega_a$. The resonators interact with themselves with coupling strength $\tilde{\lambda}$. $(b).$ A star network of resonators, each interacting with the qudit regulator via strength $\lambda$, while there is no coupling between the resonators themselves. $(c).$ The cooling protocol for a collection of resonators, $\rho_{\mathcal{V}}$, with the help of a $d$-level regulator initialized in the energy state, $\ket{k}$. The entire system is evolved through the unitary, $\hat{U}(t_{\text{opt}}^k)$, over the optimal time, $t_{\text{opt}}^k$, following which the regulator is post-selected on the eigenstate $\ket{k}$. The evolution-measurement process is repeated several times in order to bring $\rho_{\mathcal{V}}$ to its ground state, $\ketbra{\mathbf{0}}{\mathbf{0}}$, with $\ket{\mathbf{0}}$ denoting that all the resonators are in the vacuum state, $\ket{0} \otimes \ket{0} \otimes \dots$.}
\label{fig:schematic}
\end{figure*}

In the measurement-based refrigeration protocol~\cite{Zhang_PRA_2024_Fock-state_measurement, Yan_PRA_2022_cooling_measurement}, the system to be cooled comprises a set of $M$ harmonic oscillators (resonators), $\{\mathcal{V}_i\}_{i = 1}^M$, coupled to an auxiliary multilevel atomic system, i.e., a  qudit, referred to as regulator (\(R\)). Each resonator has frequency $\omega_{f_i}$ and the interaction between the systems is modeled through a Jaynes-Cummings (JC)- type interaction~\cite{Jaynes_IEEE_1963_Jaynes-Cummins, Greentree_JPB_2013_Jaynes-Cummins_50, Larson_JOSAB_2024_Jaynes-Cummins_60, Larson_IOP_2024_Jaynes-Cummins}, as illustrated schematically in Fig.~\ref{fig:schematic}. The free Hamiltonian of the composite system reads
\begin{equation}
\hat{H}_0=\sum_{k=0}^{d-1} \,\omega_k \ketbra {k}
+\sum_{j=1}^{M} \omega_{f_j}\, \hat{a}_j^\dagger \hat{a}_j ,
\label{equ:local_hamiltonian}
\end{equation}
where $\hat{a}_j$ ($\hat{a}_j^\dagger$) are the annihilation (creation) operators of the $j$-th resonator which obey the bosonic commutation relation $[\hat{a}_i, \hat{a}_j^\dagger] = \delta_{ij}$ with $\delta_{ij}$ being the Kronecker delta. For the qudit system, $\omega_k$ denotes the energy spacing between consecutive levels and $\ket{k}$ corresponds to the $k$-th energy eigenstate. To maintain symmetry in the Hamiltonian, we assume the resonant condition $\omega_{f_i} = \omega_a , \,\forall i ,$ and, for simplicity, we set $\omega_{f_i} = \omega_a = 1$, where \(\omega_k=k\omega_a\). Note that the assumption of equally spaced atomic energy levels is not essential for the entire analysis carried out in this work.

The interaction Hamiltonian, $\hat{H}_I$, depends on the network configuration and establishes correlations between the qudit and the resonator(s), which serve as the operational resource for the refrigeration process. We consider two arrangements between the resonators and the qudit - $(1)$ the linear chain, where each resonator interacts with the other ones on either side and only the first resonator interacts with the regulator (see Fig.~\ref{fig:schematic}(\(a\)) ), and $(2)$ the star configuration, where each resonator is connected to the regulator with no interaction between themselves (see Fig.~\ref{fig:schematic}(\(b\)) ). 
The respective interaction Hamiltonians corresponding to the two configurations are given by
\begin{eqnarray}
  \hat{H}_I^\text{lin} &=& \sum_{k = 0}^{d - 1} \lambda \hat{a}_1 \ketbra{k}{k-1} + \sum_{i = 1}^{M - 1} \tilde{\lambda} \hat{a}_i^\dagger \hat{a}_{i+1} + \text{h.c.} \label{eq:linear_HI} \\
\text{and}~  \hat{H}_I^{\text{star}} &=& \sum_{i = 1}^M \lambda \hat{a}_i \Big( \sum_{k = 0}^{d - 1} \ketbra{k}{k-1} \Big) + \text{h.c.} \label{eq:star_HI}
\end{eqnarray}
Here too, we set the resonator-regulator and the intra-resonator interactions, $\lambda = \tilde{\lambda} = 1$, for both network configurations.

Since the refrigeration protocol is based on a measurement-based purification scheme, it is interesting to analyze whether an initial $M$-mode mixed state of the resonators can be successfully converted to a chain of vacua. 
In particular, let us take the initial state of each resonator as the single-mode displaced squeezed thermal state,
$\rho_i(\alpha_i,z_i,\bar{n}_i) = \Big(\hat{D}(\alpha_i)\, \hat{S}(z_i)\,
\rho_{\mathrm{th}}(\bar{n}_i)\,
\hat{S}^{\dagger}(z_i)\, \hat{D}^{\dagger}(\alpha_i)\Big)$. Here, $\hat{D}(\alpha) := \exp\!\left(\alpha \hat{a}^{\dagger}-\alpha^{*}\hat{a}\right)$ is the displacement operator with parameter $\alpha = |\alpha| e^{\iota \phi}$ ($\iota = \sqrt{-1}$), and  $\hat{S}(r) := \exp\!\left[\frac{1}{2}\left(z\hat{a}^{\dagger 2}-z^*\hat{a}^{ 2}\right)\right]$ is the single-mode squeezing operator with squeezing parameter $z = r e^{\iota \theta}$, with complex conjugation being denoted by $*$. The thermal state, $\rho_{\text{th}}(\bar{n})$, characterized by the mean photon number $\bar{n}$, is given by $\rho_{\text{th}}(\bar{n}) := \sum_{m=0}^{\infty}\frac{\bar{n}^{m}}{(1+\bar{n})^{m+1}}\ketbra {m}{m}$. With the regulator initialized in the $k$-th qudit eigenstate, $\rho_R = \ketbra{k}{k}$, the global state of the system at time, $t = 0$, can be represented as  $\rho_{\text{in}}(0) = \rho_R  \otimes_{i = 1}^M \rho_i(\alpha_i,z_i,\bar{n}_i)=\rho_R \otimes \rho_{\mathcal{V}}$.

The cooling mechanism comprises the repeated realization of two steps - evolution through $\hat{U}(t) = e^{-\iota \hat{H} t}$ where $\hat{H} = \hat{H}_0 + \hat{H}_I^{\text{lin}}(\hat{H}_I^{\text{star}})$ followed by a measurement of the qudit in the basis $\{\ket{k}\}_{k = 0}^{d - 1}$ (see Fig.~\ref{fig:schematic}(\(c\)) ). If the measurement outcome is the initial state of the qudit, the procedure is continued; else, we start over from $\rho_{\text{in}}(0)$. Thus, the protocol succeeds in producing $\ket{0}^{\otimes M}$ as the final state of the resonator system only probabilistically. In order to attain the goal of near-perfect fidelity, with as few rounds as possible, the evolution time at each cycle must be carefully chosen. {This optimal time of evolution, $t_{\text{opt}}^k$, for measurement in $\ketbra{k}{k}$, is determined by the qudit dimension, the energy-level of the measurement, and the network configuration as we shall describe below.} The unnormalized density matrix describing the final state of the resonator system after $N$ rounds of evolution, conditioned on the desired measurement outcome, is given by

\begin{eqnarray}
   \nonumber \rho^N_{\mathcal{V}} &=& \Tr_R \Big(\hat{O}_k \hat{U}(t_{\text{opt}}^k) \Big)^N \rho_{\text{in}} \Big(\hat{O}_k \hat{U}(t_{\text{opt}}^k) \Big)^{\dagger N} \\
   &=& \Tr_R \Big(\hat{O}_k \hat{U}(t_{\text{opt}}^k) \hat{O}_k \Big)^N \rho_{\text{in}} \Big(\hat{O}_k \hat{U}(t_{\text{opt}}^k) \hat{O}_k \Big)^{\dagger N},
   \label{eq:unnorm_final-state}
\end{eqnarray}
where $\hat{O}_k = \ketbra{k}{k}$ denotes the projector of the $k$-th qudit eigenstate. {Note that the second equality in Eq.~\eqref{eq:unnorm_final-state} holds solely due to the fact that the qudit is initialized in $\ketbra{k}{k}$.} This further allows us to define an effective evolution operator $\hat{V}_d^k = \bra{k} \hat{U}(t_{\text{opt}}^k) \ket{k}$ for the resonator system alone, whereby we can represent the final resonator state as
\begin{eqnarray}
    \rho^N_{\mathcal{V}} &=& \frac{(\hat{V}_d^k)^N \rho_\mathcal{V} (\hat{V}_d^k)^{\dagger N}}{\Tr [(\hat{V}_d^k)^N \rho_\mathcal{V} (\hat{V}_d^k)^{\dagger N}]},
    \label{eq:final_resonator-state}
\end{eqnarray}
with the probability of success being $P_{d, N}^k = \Tr [(\hat{V}_d^k)^N \rho_\mathcal{V} (\hat{V}_d^k)^{\dagger N}]$. This effective description of the evolution of the resonator system assumes no specific interaction topology. 

In order to analyze the evolution of the resonators under the aforementioned protocol, let us first note that the total Hamiltonian conserves the number of excitations, as it commutes with the number operator, $\hat{\mathcal{N}}_e=\sum_{j=1}^{M} \hat{a}_j^\dagger \hat{a}_j+\sum_{k=0}^{d-1} k\, \ketbra{k}{k}$. {As such, the Hamiltonian, $\hat{H}$, and the unitary, $\hat{U}(t)$, are both block-diagonal in structure, with each block corresponding to a fixed excitation number. For instance, the blocks for $\langle \hat{\mathcal{N}}_e \rangle = 0, 1, 2$ are spanned by the bases $\{\ket{\mathbf{0}_\mathcal{V},0_R}\}, \{\ket{\mathbf{1}_\mathcal{V}, 0_R}, \ket{\mathbf{0}_\mathcal{V}, 1_R}\}, \{\ket{\mathbf{2}_\mathcal{V}, 0_R},\ket{\mathbf{1}_\mathcal{V}, 1_R}$, $\ket{\mathbf{0}_\mathcal{V}, 2_R}\}$ respectively.} Here, by $\ket{\mathbf{n}}$, we represent the global energy-eigenstate of the free resonator Hamiltonian with energy $n$, i.e., the total number of excitations in all the oscillators together is $n$. {Hence, the effective evolution operator is block-diagonal in the resonator energy eigenbasis}

\begin{eqnarray}
    \hat{V}_d^k = \sum_{i = 0}^{\infty} \lambda_{i,d}^k \ketbra{\mathbf{i}}{\mathbf{i}},
    \label{eq:effective_operator}
\end{eqnarray}
while the probability of success for the protocol reads

\begin{eqnarray}
    P_{d, N}^k = \sum_{i = 0}^{\infty} |\lambda_{i,d}^k|^{2N} (\rho_{\mathcal{V}})_{ii},
    \label{eq:probability}
\end{eqnarray}
where \((\rho_{\mathcal{V}})_{ii}\) denotes matrix element \((i,i)\) of \(\rho_{\mathcal{V}}\).
Finally, the fidelity of $\rho^N_{\mathcal{V}}$ with the state $\ket{\mathbf{0}} = \ket{0}^{\otimes M}$ is 

\begin{eqnarray}
    F_{d,N}^k = |\bra{\mathbf{0}} \rho_\mathcal{V}^N \ket{\mathbf{0}}| = \frac{1}{P_{d,N}^k} |\lambda_{0, d}^k|^{2N} (\rho_{\mathcal{V}})_{00}.
    \label{eq:fidelity}
\end{eqnarray}
The fidelity thus depends on the evolution dynamics through $\lambda_{0,d}^k$ and attains its maximum when $|\lambda_{0, d}^k|^{2N} \to 1$. {Since $|\lambda_{i, d}^k| \leq 1 ~\forall~ i$, the condition for optimality is given as $|\lambda_{0, d}^k| = 1$ from which we can derive the optimal evolution time, $t_{\text{opt}}^k$.} Note that the optimality condition not only ensures near-perfect fidelity, but also does so with the minimum number of evolution-measurement rounds.

\section{Dimensional advantage in single oscillator cooling using qudit regulators: a no-go theorem for qubits}
\label{sec:single_oscilator_cooling}

Let us first concentrate on the dynamics of a single oscillator coupled to a single qudit. A system comprising a single oscillator and a qudit has also been used to design a photon-statistics analyser~\cite{Schuster_Nature_2007_photon-statistics} and perform qubit-photon conditional logic~\cite{Vlastakis_Science_2013_qubit-photon_conditional-logic}, as well as to prepare highly squeezed Gaussian and non-Gaussian GKP states~\cite{Eickbusch_NP_2022_SNAP-gate}. Further, the following treatment would allow us to gain insight into $t_{\text{opt}}^k$ for different qudit dimensions, especially to determine the optimality conditions for a generic network configuration with an arbitrary number of resonators. For a single resonator, both the linear and star frameworks are represented by the Hamiltonian, $\hat{H} = \hat{a}^\dagger \hat{a} + \sum_{k = 0}^{d - 1} k \ketbra{k}{k} + \sum_{k = 0}^{d - 2} \Big( \hat{a}^\dagger \ketbra{k}{k + 1} + \hat{a} \ketbra{k + 1}{k} \Big)$.

\subsection{Qubit as the regulator: no-cooling theorem }
When the regulator is two dimensional, there are only two measurement options - $\hat{O}_0 = \ketbra{0}{0}$, and $\hat{O}_1 = \ketbra{1}{1}$, and the respective evolution operators, $\hat{V}_2^{0/1}$, have a block corresponding to $\langle \hat{\mathcal{N}}_e\rangle = 0$ spanned by $\{\ket{0_\mathcal{V}, 0_R}\}$, while the remaining blocks for $\langle\hat{\mathcal{N}}_e\rangle \geq 1$ are spanned by $\{\ket{(i+1)_{\mathcal{V}}, 0_R}, \ket{i_\mathcal{V}, 1_R}\}_{i = 0}^\infty$. The effective evolution operators for measurement in the ground and excited-states of the qubit are, respectively, given by

\begin{eqnarray}
    \hat{V}_2^0 &=& \sum_{i = 0}^{\infty} e^{- \iota i t} \cos \sqrt{i} t \ketbra{i}{i}, \label{eq:V0_qubit}\\
 \text{and}~   \hat{V}_2^1 &=& \sum_{i = 0}^{\infty} e^{-\iota (i + 1) t} \cos \sqrt{i+1} t \ketbra{i}{i}. \label{eq:V1_qubit}
\end{eqnarray}
We can immediately infer that $|\lambda_{0,2}^0| = 1 ~\forall~ t$, while $|\lambda_{0,2}^1| = 1 \implies t_{\text{opt}}^1 = p \pi (p = 0, 1, \dots)$. Even though it is not possible to determine the optimal time for measurement in $\ketbra{0}{0}$ from the structure of $\hat{V}_2^0$, we can do so by analyzing the probability expression in Eq.~\eqref{eq:probability} with $d = 2, k = 0$ as

\begin{eqnarray}
    P_{2, N}^0 = \sum_{i=0}^{\infty} \cos^{2N}\!\left(\sqrt{i}\, t\right) (\rho_{\mathcal{V}})_{ii}.
    \label{eq:prob_qubit_0}
\end{eqnarray}
The dominant contribution to $P_{2,N}^{0}$ arises from $(\rho_{\mathcal{V}})_{00}$, followed by $(\rho_{\mathcal{V}})_{11}$, since in most physical systems the higher energy levels are less populated for moderate energies. Suppressing the contribution from $(\rho_{\mathcal{V}})_{11}$ would significantly enhance the fidelity given by Eq.~\eqref{eq:fidelity} (since it contains $P_{2,N}^{0}$ in the denominator) while allowing convergence to $\ket{0}_{\mathcal{V}}$ with a small number of measurements. This is achieved by choosing the optimal interaction time $t_{\text{opt}}^1=\frac{\pi}{2}$, for which the $i=1$ term vanishes in the probability expression. With these choices for $t_{\text{opt}}^k$, the fidelities of obtaining the resonator in the vacuum state after $N$ rounds of evolution and measurement in $\ketbra{0}{0} ~\text{and}~ \ketbra{1}{1}$ are respectively given by

\begin{eqnarray}
F_{2,N}^{0} &=& \frac{(\rho_\mathcal{V})_{00}}{C_{00}+\sum_{i=1}^{\infty}\cos^{2N}\!\left(\sqrt{i}\frac{\pi}{2}\right) (\rho_\mathcal{V})_{ii}}, \label{eq:fidelity_qubit_0}  \\
~\text{and}~ F_{2,N}^{1} &=& \frac{(\rho_\mathcal{V})_{00}}{C_{00}+\sum_{i=1}^{\infty} \cos^{2N}\!\left(\sqrt{i+1}\,\pi\right) (\rho_\mathcal{V})_{ii}}\label{eq:fidelity_qubit_1},
~~~~
\end{eqnarray}
with the denominators representing the corresponding probabilities, $P_{2, N}^{0/1}$. We are now in a position to present the first crucial result of this work, that cooling oscillators with qubit auxiliary states is not possible with unit fidelity.

\THEO{$\mathbf{1}$} \textbf{No-cooling with qubits:} \emph{It is impossible to cool a resonator with unit fidelity at a non-vanishing probability using the repetitive evolution and measurement technique when the auxiliary regulator is a qubit.}

\begin{proof}
Consider a general initial state of the resonator written in the Fock basis as
\begin{equation}
\rho_\mathcal{V}=\sum_{m,n = 0}^{\infty} C_{mn}\ketbra{m}{n},
\label{eq:rho1_0}
\end{equation}
where $C_{mn}$ are the matrix elements of the density operator satisfying $\sum_{m = 0}^\infty C_{mm} = 1$. From Eqs.~\eqref{eq:fidelity_qubit_0} and ~\eqref{eq:fidelity_qubit_1}, {we note that $F_{2,N}^{i} \times P_{2, N}^i = (\rho_{\mathcal{V}})_{00} = C_{00}$, whence $F_{2, N}^{i} = 1 \implies P_{2, N}^{i} = C_{00} ~\forall~ i \in \{0, 1\}$.} Therefore, for perfect fidelity of obtaining a vacuum state after a large number of evolutions through $t_{\text{opt}}^k$ followed by measurement on $\ketbra{k}{k}$, we must have $\cos^{2N} \sqrt{i} \pi/2 \to 0$ or $\cos^{2N} \sqrt{i+1} \pi \to 0$ as $N \to \infty$, which requires the cosine functions to be less than unity. However, non-vanishing contributions arise when $\cos\!\left(\sqrt{i}\pi/2 \right) = \cos\!\left(\sqrt{i+1}\pi\right) = \pm1,$ which are satisfied if
\begin{equation}
i = \begin{cases}
    4 n^2 ~\text{for measurement in} \ketbra{0}{0} \\
    n^2-1 ~\text{for measurement in} \ketbra{1}{1},
\end{cases}
\label{eq:no-go_proof_1}
\end{equation}
for integer $n\geq1$. Consequently, the fidelities asymptotically reduce to
\begin{eqnarray}
F_{2,N}^{0} &=& \frac{C_{00}}{\sum_{i=0,4,16,\ldots} C_{ii}}, \label{eq:no-go_proof_2} \\
\text{and}~ F_{2, N}^{1} &=& \frac{C_{00}}{C_{00}+\sum_{i=1}^{\infty} \cos^{2N}\!\left(\sqrt{i+1}\,\pi\right)C_{ii}}. \label{eq:no-go_proof_3}
\end{eqnarray}

For physically relevant initial states, the populations, $C_{ii}$, decay rapidly with increasing $i$, and contributions beyond $i \geq 10$ can be neglected, yielding
\begin{eqnarray}
F_{2,N}^{0} &\approx& \frac{C_{00}}{C_{00}+C_{44}} = 1-\frac{C_{44}}{C_{00}+C_{44}} < 1, \\
F^1_{2, N} &\approx& \frac{C_{00}}{C_{00}+C_{33}+C_{88}} = 1-\frac{C_{33}+C_{88}}{C_{00}+C_{33}+C_{88}} < 1. ~~~~~~~~
\end{eqnarray}
Hence, irrespective of the desired measurement outcome, a qubit auxiliary system cannot cool a single resonator to its ground state with unit fidelity and nonvanishing success probability. This completes the proof. 
\end{proof}

\subsection{Overcoming the limitation by increasing dimension}
The question now arises whether increasing the dimension of the auxiliary system can resolve the above problem and allow for perfect cooling of the resonator ($F_{d, N}^k = 1$) with a non-vanishing probability.

\THEO{$\mathbf{2}$} \emph{When the regulator is three-dimensional, the resonator can be cooled to a ground state with unit fidelity and non-vanishing probability.}

\begin{proof}
Let us concentrate on the case for $d = 3$, in which case, the effective evolution operators for measurement in $\{\ketbra{k}{k}\}_{k = 0}^{^2}$ read

\begin{eqnarray}
   \nonumber \hat{V}_3^0 &=& \ketbra{0}{0} + \cos t \ketbra{1}{1} + \\
   &&\sum_{i = 0}^\infty \frac{1 + i + (2 + i) \cos \sqrt{3 + 2 i} t}{3 + 2 i} \ketbra{i + 2}{i + 2}, ~~~~~~ \label{eq:qutrit_effective_0} \\
   \hat{V}_3^1 &=& \sum_{i = 0}^\infty \cos \sqrt{2i + 1} t \ketbra{i}{i}, \label{eq:qutrit_effective_1} \\\text{and}~
   \hat{V}_3^2 &=& \sum_{i = 0}^\infty \frac{2 + i + (1 + i) \cos \sqrt{3 + 2 i} t}{3 + 2 i} \ketbra{i}{i}. \label{eq:qutrit_effective_2}
\end{eqnarray}
Corresponding to the three measurement choices, $\{\ketbra{k}{k}\}_{k = 0}^2$, the respective optimal evolution times are given by {$t_{\text{opt}}^k = \pi/2, \pi, 2\pi/\sqrt{3}$.} For the same generic initial state as considered in the proof of Theorem $1$, the probability expressions assume the form
\begin{eqnarray}
  \nonumber  P_{3, N}^k = C_{00} + \sum_{i = 0}^\infty f_i(t_{\text{opt}}^k)^{2N} \ketbra{i + (2 - k)}{i + (2 - k)}, \\
    \label{eq:prob_qutrit}
\end{eqnarray}
where $f_i(t_{\text{opt}}^k)$ represents the coefficients inside the summations in Eqs.~\eqref{eq:qutrit_effective_0} - ~\eqref{eq:qutrit_effective_2} at $t = t_{\text{opt}}^k$ for $k = 0, 1, 2$. Similar to the qubit case, unit fidelity implies $P_{3, N}^k = C_{00}$, whence we must have $|f_i(t_{\text{opt}}^k)| < 1 ~\forall~ i \implies f_i(t_{\text{opt}}^k)^{2N} \to 0 ~\text{for}~ N \to \infty$. On the other hand, $|f_i(t_{\text{opt}}^k)| = 1$ is achieved only if
\begin{eqnarray}
    i = \begin{cases}
        \frac{16 n^2 - 3}{2} ~\text{for}~ k = 0 \\
        \frac{n^2 - 1}{2} ~\text{for}~ k = 1 \\
        \frac{3n^2 - 12}{8} ~\text{for}~ k = 2.
    \end{cases}
    \label{eq:qutrit_optimal-conditions}
\end{eqnarray}
We observe that, except for $k = 1$, no integer value of $n$ yields a feasible integer value for $i$. This implies that for a sufficiently large number of evolution-measurement cycles, it will be possible to attain unit fidelity when measuring the auxiliary qutrit in $\ketbra{0}{0}$ or $\ketbra{2}{2}$. 
\end{proof}
{\it Therefore, dimensional advantage is apparent when using $d = 3$ auxiliary states, in the sense that perfect cooling of the resonator with a non-vanishing probability is possible.}

\subsection{Quantifying the dimensional advantage of cooling with higher-dimensional regulators}
\label{subsec:dim-adv_single}

We now elucidate the dimensional advantage obtained in the cooling protocol for a single resonator by increasing the dimension of the regulator. There are two principal aspects in which the dimensional advantage is prominent: $\mathbf{(1)}$ the number of cycles, $N$, required to achieve perfect cooling decreases with $d$, and $\mathbf{(2)}$ increasing the dimension of the auxiliary qudit allows for cooling systems with higher energies. Let us restrict the maximum number of evolution-measurement rounds to $N_{\max} = 100$. This would allow us to make a fair comparison between regulators of different dimensions, since with a sufficiently large number of rounds, it would be theoretically possible to obtain the resonator in a final vacuum state. However, an arbitrary number of rounds, albeit guaranteeing the production of $\ket{0}$ with near-perfect fidelity, would eventually render the probability of success vanishingly small.

\begin{table}[t]
\centering
\begin{tabular}{|c|c|c|c|c|c|c|c|}
\hline
$k$&$0$&$1$&$2$&$3$&$4$&$5$&$6$\\
\hline
$t_{\text{opt}}^k$&$\frac{\pi}{2} \approx 1.57$&$\pi \approx 3.14$&$\frac{2 \pi}{\sqrt{3}} \approx 4.44$&$59.25$&$88.05$&$157.95$&$217.71$\\\hline
\end{tabular}
\caption{\justifying The optimal time of evolution, $t_{\text{opt}}^k$, when the regulator is measured in the level $k$. All values in the second row are correct up to the second decimal place.}
\label{tab:optimal_t}
\end{table}


With this protocol structure in mind, let us first determine the optimal time required when the regulator is prepared and measured at different levels. As has already been discussed beforehand, $t_{\text{opt}}^k = \pi/2, \pi, 2\pi/\sqrt{3}$ for $k = 1, 2, 3$ corresponding to measurement in the ground, excited, and first excited-states. Our analysis suggests that with increasing $d$, the solution of $|\lambda_{0, d}^{d - 1}| \approx 1$ comprises $d$ cosine terms whose arguments are the roots of the probabilists' Hermite polynomial of order $d$, $He_d(x) = (-1)^d e^{x^2/2} \frac{d^d}{dx^d} e^{-x^2/2}$, and studying the effective evolution operators, $\hat{V}_{d > 3}^{k > 2}$, becomes analytically intractable, and we must resort to numerical techniques. By constructing the Hamiltonian for single oscillator cooling in an effective subspace of dimension $~50$, and finding the solution to $|\lambda_{0, d}^k| \approx 1$, we observe that the optimal evolution time for one cycle increases with $k$ (see Table.~\ref{tab:optimal_t}). The increment of the optimal time with $k \in \{0, 1, 2, 3\}$ has a happy consequence that, in practical setups, it is not necessary to perform the repeated measurements in very quick succession, allowing for better control. However, the considerably larger values of $t_{\text{opt}}^{k \geq 4}$ may be detrimental since the system has enough time to decohere due to environmental interactions during the evolution. 

\begin{table}[t]
\centering
\begin{tabular}{|c|c|c|c|c|c|c|c|c|c|}
\hline
$d$ & \multicolumn{3}{c|}{$N$}     & \multicolumn{3}{c|}{$F_{d, N}^k$} & \multicolumn{3}{c|}{$P_{d, N}^k$} \\
\hline
    & $k = 0$ & $k = 1$ & $k = 2$ & $k = 0$   & $k = 1$  & $k = 2$  & $k = 0$   & $k = 1$   & $k = 2$  \\
\hline
$2$ & $100$   & $100$   &   -      & $0.981$   & $0.953$  &     -     & $0.656$   & $0.675$   & $0.643$  \\
$3$ & $61$    & $100$   & $38$    & $0.999$   & $0.981$  & $0.999$  & $0.643$   & $0.656$   & $0.643$  \\
$4$ & $2$     & $21$    & $13$    & $0.999$   & $0.999$  & $0.999$  & $0.643$   & $0.643$   & $0.643$  \\
$5$ & $2$     & $14$    & $30$    & $0.999$   & $0.999$  & $0.999$  & $0.643$   & $0.643$   & $0.643$  \\
$6$ & $2$     & $13$    & $13$    & $0.999$   & $0.999$  & $0.999$  & $0.643$   & $0.643$   & $0.643$ \\
\hline
\end{tabular}
\caption{\justifying The number of evolution-measurement cycles, $N$, required to obtain a near-perfect fidelity, i.e., $F_{d, N}^k \to 1$ with the vacuum, along with the corresponding success probability, $P_{d, N}^k$, when a single resonator in an initial displaced squeezed thermal state, $\rho_{\mathcal{V}} = \rho(\alpha = 0.4, r = 0.1, \bar{n} = 0.4)$, is coupled to a qudit of dimension $d$ and measurements are performed on the energy level $k$. All the data for $F_{d, N}^k$ and $P_{d, N}^k$ are correct up to the third decimal place.}
\label{tab:N-fid-prob-012}
\end{table}

We now proceed to study the benefit of increasing the regulator dimensions, in terms of the number of cycles required, for the cooling protocol using the optimal evolution time for each cycle. We restrict the measurement levels to $k \leq 2$. For this purpose, we first consider a displaced squeezed thermal state with $\alpha=0.4$, $r=0.1$, and $\bar{n}=0.4$, which lie well within experimentally accessible regimes~\cite{Vahlbruch_PRL_2016_15dB-squeezing}. The qudit regulator is initialized in the state, $\ket{k}$, which we want to postselect on at every measurement step. We observe that, for measurement on the energy level $\ket{k}$, the number of cycles, $N$, required to obtain perfect fidelity decreases rapidly with $d$ (see Table.~\ref{tab:N-fid-prob-012}). This advantage is particularly pronounced for measurement on $\ketbra{0}{0}$ and $\ketbra{1}{1}$. Furthermore, consistent with Theorem $1$, a qubit regulator ($d = 2$) can never help to achieve perfect cooling, with the maximum fidelity reaching $F_{2, 100}^0 = 0.981$. Similarly, when a regulator with $d = 3$ is measured in the excited-state, $k = 1$, near-perfect fidelity cannot be achieved as argued for the case of qutrits in the previous section. Note that, although we restrict the maximum number of cycles to $N_{\max} = 100$, the fidelity cannot be improved further even for $N \to \infty$ since it saturates to the aforementioned values. On the other hand, for $d \geq 4$, the fidelity always reaches close to unity irrespective of the qudit level on which the measurement is performed. Once the fidelity saturates to $F_{d, N\leq 100}^{k<=2} \approx 0.999$, the probability also assumes a constant value, $P_{d, N<=100}^{k<=2} \approx 0.643$ in this example, as is expected from the condition $F_{d, N}^k \times P_{d, N}^k = (\rho_\mathcal{V})_{00}$. 

\begin{figure}[ht]
\includegraphics[width=\linewidth, height=5.9cm]{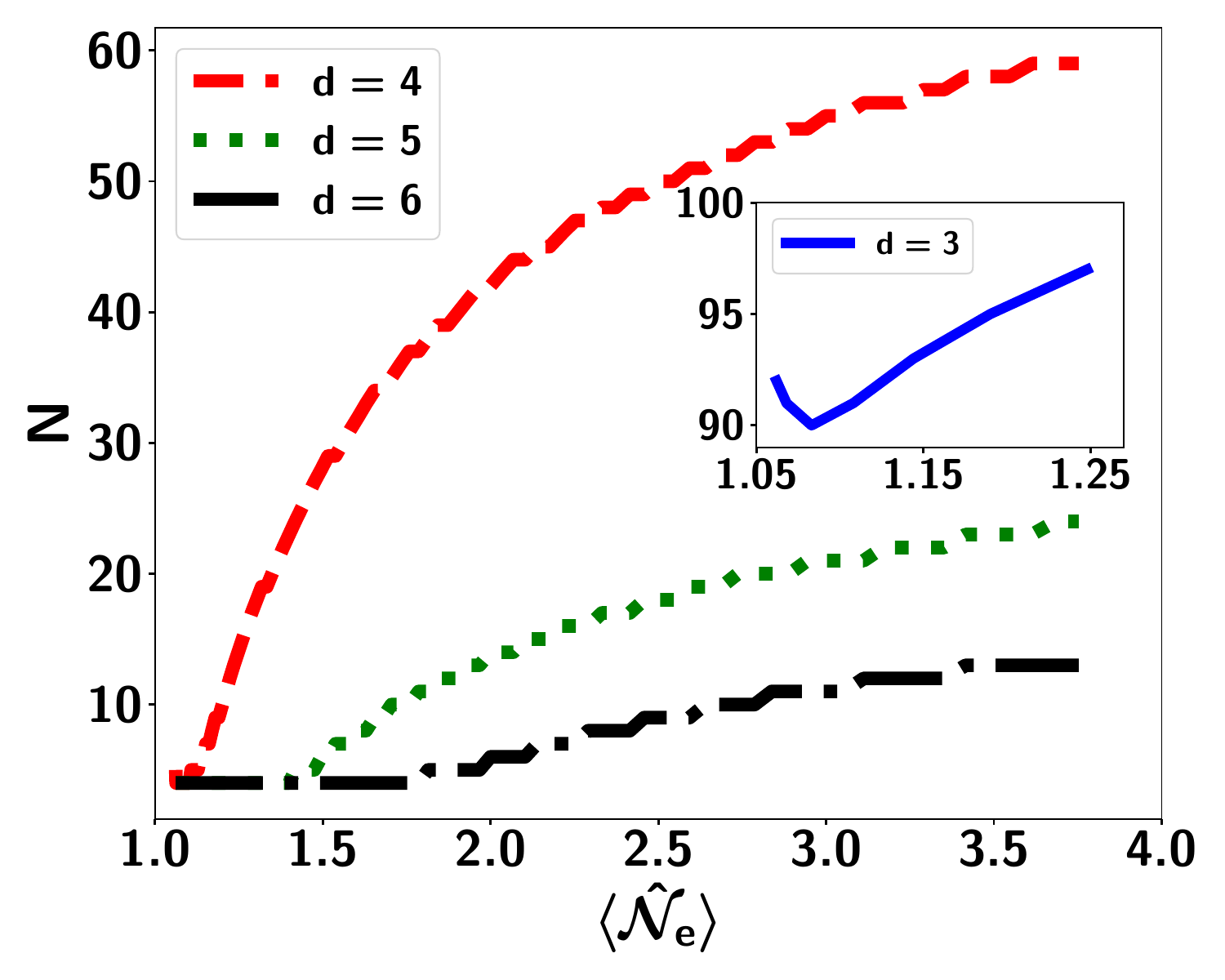}
\captionsetup{justification=Justified,singlelinecheck=false}
\caption{\textbf{Energy of the resonator that admits near-perfect cooling with qudit auxiliary regulators.} Required number of evolution-measurement cycles, $N$ (ordinate), against the average resonator energy, $\langle \hat{\mathcal{N}}_e \rangle$ (abscissa), for different regulator dimensions, $d = 4$ (red dashed line), $d = 5$ (green dotted line), and $d = 6$ (black dash-dotted line). The inset shows the same for $d = 3$ (blue solid line), for which $N$ is considerably higher even for a low-energetic resonator. Both axes are dimensionless.}
\label{fig:measure_energy} 
\end{figure}

The dimensional advantage on the number of required cycles can also be studied with respect to the energy of the resonator to be cooled. In particular, we demonstrate in Fig.~\ref{fig:measure_energy} that to obtain $F_{d, N\leq 100}^0 \geq 0.999$ with a fixed regulator dimension, $N$ increases with the average energy, $\langle \hat{\mathcal{N}_e} \rangle$, of the resonator, when measurement is performed on the ground state. This is because, as the energy increases, the population of the vacuum, $(\rho_\mathcal{V})_{00}$, in the initial resonator state decreases, and it becomes more difficult to bring the system down to its ground state. This effect is most pronounced for $d = 3$, which cannot cool resonators having energy $\langle \hat{\mathcal{N}_e} \rangle > 1.261$ with near-perfect fidelity, when the total number of rounds is restricted. On the other hand, for a fixed resonator energy, the number of required cycles of evolution and measurement decreases steadily with $d$, thereby again highlighting the {\it dimensional advantage} (see Fig.~\ref{fig:measure_energy}). 

The favorable impact of increasing the regulator dimension observed above motivates us to estimate the maximum energy of the initial resonator that allows near-perfect cooling with $N_{\max} = 100$ with a regulator of a fixed dimension. For this purpose, we set our target fidelity to $F_{d, N}^k \geq 0.999$ and demand a success probability $P_{d, N}^k \geq 0.1$, i.e., we try to find an upper bound on the initial resonator energy such that it can be cooled with near perfect fidelity and a moderate success probability. We observe that with an increase in the regulator dimension, the maximum energy for successful cooling also increases. For repeated conditional measurements on $\ketbra{0}{0}$ and $\ketbra{1}{1}$, cooling with the aforementioned conditions is possible up to $\langle \hat{\mathcal{N}_e} \rangle_{\max} = 11.636$ for $d \geq 4$.  Thus, increasing the regulator dimension further cannot aid in cooling systems of higher energies. This saturation in the maximum energy may be an artefact of restricting the number of measurements to $N_{\max} = 100$. For measurement on $\ketbra{2}{2}$, the energy threshold increases gradually as $\langle \hat{\mathcal{N}_e} \rangle_{\max} = 1.333, 9.737, 10.105, 11.636$ for $d = 3, 4, 5, 6$ respectively. This suggests that measuring in the ground or excited-state is more favorable for successful cooling, independent of the increase in dimension. Furthermore, the number of cycles required to achieve the conditions on the fidelity and probability delineated above also decreases with $d$: for $\ketbra{0}{0}$ measurements, we require $N_{\max} = 100, 51, 43, 25$ for $d = 3, 4, 5, 6$ respectively, while for conditioning on $\ketbra{1}{1}$, the respective maximum number of cycles become $N_{\max} = 100, 86, 67, 40$. Thus, the dimensional advantage is again reflected in the decrement of the number of required evolution-measurement rounds. 

\subsubsection*{Auxiliary oscillator system as the regulator}

After establishing the dimensional benefit of qudit regulators, it will be interesting to ask -- ``instead of a finite-dimensional system, if the regulator is another harmonic oscillator, can we reduce \(N\) further?''
We shall restrict our analysis to the Gaussian regime so that phase-space techniques comprising the first and second moments of the system alone may be employed to study the dynamics of cooling. We recall that any Gaussian system can be fully characterized by its displacement vector (first moments of the state) and its covariance matrix (second moments), and refer the reader to Appendix~\ref{sec:app1} for a primer on Gaussian CV systems. 
As before, the resonator is chosen to be a displaced squeezed thermal state, $\rho_\mathcal{V} = \rho(\alpha, r, \bar{n})$, while the regulator is initialized in vacuum, $\ketbra{0}{0}_R$, such that the respective displacement vectors and covariance matrices read

\begin{eqnarray}
   && \mathbf{d}_\mathcal{V} = (\alpha_1, \alpha_2)^T;~~ \mathbf{d}_R = (0, 0)^T ~~ \text{and} \label{eq:disp_initial} \\
   && \sigma_\mathcal{V} = (\bar{n} + 1/2) \begin{pmatrix}
    e^{2 r} & 0 \\
    0 & e^{-2r}
    \end{pmatrix}; ~~ \sigma_R = \begin{pmatrix}
    \frac{1}{2} & 0 \\
    0 & \frac{1}{2}
    \end{pmatrix}. \label{eq:cov_initial}
\end{eqnarray}
The joint system, $\rho_{\text{in}} = \rho_\mathcal{V} \otimes \ketbra{0}{0}_R$, is then defined by the direct sum of the respective first and second moments, i.e., $\mathbf{d}_{\text{in}} = \mathbf{d}_\mathcal{V} \oplus \mathbf{d}_R$ and $\sigma_{\text{in}} = \sigma_\mathcal{V} \oplus \sigma_R$. The symplectic transformation corresponding to evolution through the total Hamiltonian, $\hat{H} = \hat{a}_{\mathcal{V}}^\dagger \hat{a}_{\mathcal{V}} + \hat{a}_{R}^\dagger \hat{a}_{R} + \hat{a}_{\mathcal{V}}^\dagger \hat{a}_{R} + \hat{a}_{R}^\dagger \hat{a}_{\mathcal{V}}$ is given by $ S_H = \begin{pmatrix}
        \mathcal{A} & \mathcal{B} \\
        \mathcal{B} & \mathcal{A}
    \end{pmatrix}$, where

\begin{eqnarray}
   && \nonumber \mathcal{A} = \begin{pmatrix}
        \cos^2 t & \cos t \sin t \\
        - \cos t \sin t & \cos^2 t
    \end{pmatrix}, \\
   && \mathcal{B} = \begin{pmatrix}
        -\sin^2 t & \cos t \sin t \\
        \cos t \sin t & -\sin^2 t
    \end{pmatrix}. \label{eq:symplectic_H}
\end{eqnarray}
Here $t$ is the time for each round of evolution. 

\THEO{$\mathbf{3}$}~\emph{Taking a harmonic oscillator as a regulator, only a single evolution followed by a Gaussian measurement is enough to cool the oscillator to its ground state with a unit fidelity and non-zero probability. }

\begin{proof}
The final joint system after one evolution is characterized by $\mathbf{d}^1 = S_H \mathbf{d}_{\text{in}}$ and $\sigma^1 = S_H \sigma_{\text{in}} S_H^T$ (see Appendix~\ref{sec:app1} for the symplectic picture). Note that the joint total output displacement vector and covariance matrix can be decomposed in terms of the corresponding output terms of the resonator and regulator as
\begin{eqnarray}
   && \nonumber \mathbf{d}^1 = (\mathbf{d}^1_{\mathcal{V}}, \mathbf{d}^1_R)^T ~~\text{and} \\
   && \sigma^1 = \begin{pmatrix}
       \sigma_\mathcal{V}^1 & \sigma_{\mathcal{V} R}^1 \\
       \sigma_{\mathcal{V} R}^{1T} & \sigma_\mathcal{R}^1
   \end{pmatrix}, \label{eq:disp-cov_output}
\end{eqnarray}
where $\mathbf{d}(\sigma)_{\mathcal{V}(R)}^1$ represents the displacement vector (covariance matrix) of the resonator (regulator) after one evolution over time $t$, while $\sigma_{\mathcal{V} R}^1$ encompasses the correlations created between the two subsystems. The measurement is further taken to be Gaussian as well, and is conditioned on the vacuum state represented by $\mathbf{d}_{0} = (0, 0)^T$ and $\sigma_0 = \text{diag}(1/2, 1/2)$. As such, the post-measurement state of the resonator has a displacement vector given by

\begin{eqnarray}
  \nonumber && (\mathbf{d}^1_{\mathcal{V}})_{\text{out}} = \mathbf{d}^1_{\mathcal{V}} + \sigma^1_{\mathcal{V} R} \frac{1}{(\sigma^1_R + \sigma_0)} (\mathbf{d}_0 - \mathbf{d}^1_R) \\
  &=& 4 \cos t \Big(f_+ (\alpha_1, \alpha_2, r, \bar{n}, t) , f_- (\alpha_1, \alpha_2, r, \bar{n}, t) \Big)^T, \label{eq:final_disp_oscillator}
\end{eqnarray}
where, $f_{\pm}(\alpha_1, \alpha_2, r, \bar{n}, t) = e^{2r} \alpha_2 \sin t/\Big(e^{2r} (3 + \cos 2 t) + 2(1 + 2 \bar{n}) \sin^2 t \Big) \pm \alpha_1 \cos t/\Big((3 + \cos 2 t) + 2 e^{2r}  (1 + 2 \bar{n}) \sin^2 t \Big)$. Evidently, choosing $t_{\text{opt}}^0 = \pi/2$ results in $(\mathbf{d}^1_{\mathcal{V}})_{\text{out}} = \mathbf{d}_0 = (0, 0)^T$, the vacuum displacement vector. Furthermore, at $t = t_{\text{opt}}^0 = \pi/2$, the final covariance matrix of the resonator reads

\begin{eqnarray}
  \nonumber  (\sigma^1_\mathcal{V})_{\text{out}} &=& \sigma^1_\mathcal{V} - \sigma^1_{\mathcal{V} R} \frac{1}{(\sigma^1_R + \sigma_0)} \sigma^{1T}_{\mathcal{V} R} \\
  &=& \begin{pmatrix}
      1/2 & 0 \\
      0 & 1/2
  \end{pmatrix}, \label{eq:final_cov_oscillator}
\end{eqnarray}
which is again the covariance matrix of the vacuum.
The corresponding probability, $P_{d = \infty, 1}^0$ depends on the initial state parameters and still obeys the condition of $F_{d = \infty, 1}^0 \times P_{d = \infty, 1}^0 = (\rho_{\mathcal{V}})_{00}$ with

\begin{eqnarray}
    P_{d = \infty, 1}^0 = \frac{\exp \Big( - \frac{2 \alpha_1^2}{1 + e^{2r}(1 + 2 \bar{n})} - \frac{2 e^{2r} \alpha_2^2}{1 + e^{2r} + 2 \bar{n}} \Big)}{\pi \sqrt{(1 + \bar{n})^2 \cosh^2 r - \bar{n}^2 \sinh^2 r}}, \label{eq:prob_oscillator}
\end{eqnarray}
which is non-zero.
\end{proof}
{\it Therefore, in order to perfectly cool a resonator using another one as the regulator requires only one round of evolution and conditional measurement, thereby attaining the highest level of dimensional advantage in terms of the number of cycles necessary.}

\section{Simultaneous cooling of multiple resonators for different network configurations}
\label{sec:multi_oscillator_cooling}

We are now in a position to study the cooling of a network of resonators coupled to a single regulator. For this purpose, we consider two distinct interaction topologies - linear network (Fig.~\ref{fig:schematic}(\(a\)) ) and star network (Fig.~\ref{fig:schematic}(\(b\)) ). While the linear network has applications as a quantum repeater chain~\cite{Li_AQT_2023_repeater}, the star network may be used for Wigner tomography of multimode CV states using displaced parity measurements on the auxiliary system~\cite{Wang_Science_2016_Schroedinger-cat_two-boxes}. 

Deriving an analytical expression for the effective evolution operator, $V_d^i(t)$, and subsequently the optimal evolution time, $t_{\text{opt}}^k$, that guarantees all $M$ resonators have their final state as vacuum is generally intractable because {the constant-energy blocks beyond $\langle \hat{\mathcal{N}}_e \rangle = 0$ }grow in size rapidly for increasing $M$. This is because for $\langle \hat{\mathcal{N}}_e \rangle \geq 1$, multiple resonators allow for several eigenstates which can span the subspace. Therefore, we investigate the cooling performance numerically by constraining our measurement level at $\ketbra{0}{0}$ for regulators of any dimension after evolution through $t_{\text{opt}}^0 = \pi/2$. The initial state of every resonator is taken to be a displaced squeezed thermal state with $\alpha = 0.5$, $r = 0.2, \theta = 0$, and $\bar{n} = 0.5$, and the maximum number of evolution-measurement cycles is again restricted to $N_{\max} = 100$.

\begin{figure}[ht]
\includegraphics[width=\linewidth]{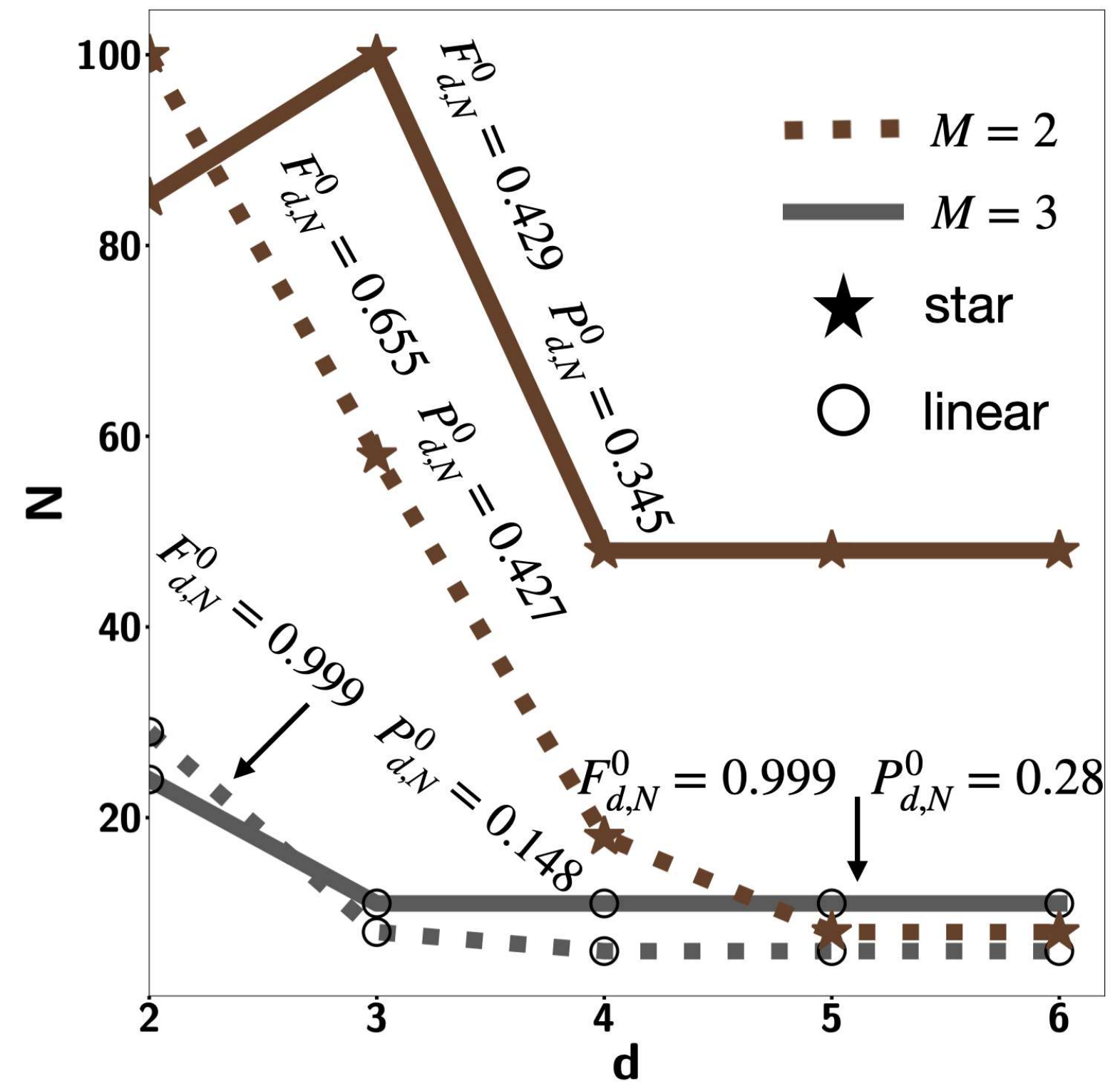}
\captionsetup{justification=Justified,singlelinecheck=false}
\caption{{\bf Minimum number of cycles required for cooling different network sizes and topologies:} Number of evolution--measurement rounds, $N$ (ordinate), as a function of the regulator dimension, $d$ (abscissa), for cooling a network of $M=2$ (dotted line) and $M=3$ (solid line) oscillators in the linear (hollow circles) and star (solid stars) interaction topologies. Increasing the regulator dimension produces a clear dimensional advantage, reducing the required number of rounds and leading to saturation. The linear network enables near-perfect cooling with substantially fewer rounds, while the star configuration saturates at lower fidelities but with a higher success probability. The values cited are correct up to the third decimal place, and both the axes are dimensionless.}
\label{fig:number_vs_dimention_star_linear} 
\end{figure}

For both star and linear networks dimensional advantage associated with increasing $d$ is apparent from a reduction in the number of rounds needed to saturate the fidelity and probability to their best-possible values as dictated by $F_{d, N}^0 \times P_{d, N}^0 = (\rho_{\mathcal{V}})_{00} = \prod_{i = 1}^M  (\rho_i)_{00}$ (see Fig. \ref{fig:number_vs_dimention_star_linear}). In both configurations, regulators with $d>3$ substantially reduce $N$ until it too saturates to a particular value. The numerical convergence of the fidelities are considered within an accuracy of $\mathcal{O}(10^{-3})$.

Although both network topologies exhibit dimensional advantage, there are intricate differences between the two in terms of the maximum achievable fidelity and the number of required rounds of evolution and conditional measurement as shown in Fig. \ref{fig:number_vs_dimention_star_linear}:
$(1)$ The linear network facilitates perfect cooling of all the resonators for all considered qudit regulators. This may be attributed to the fact that each oscillator is connected to others, which allows for better exchange of excitations between them, finally leading to optimal cooling of the entire network. In contrast, for the star network, the fidelities saturate at lower values, namely $F_{d,N}^0 \simeq 0.655$ ($M = 2$) and $F_{d,N}^0 \simeq 0.429$ ($M = 3$) for $d > 3$. Note that increasing the regulator dimension initially increases the fidelity in the star configuration before attaining the saturation value; $(2)$ In terms of $N$, we further observe that the linear network admits perfect cooling with a significantly lower number of rounds than the star counterpart, case in point being $N = 18$ and $N = 45$ for the linear and star networks, respectively, for $M = 3$; (\(3\)) The saturation in $N$ occurs at much lower regulator dimension, $d = 4$, for the linear case as compared to $d = 6$ for the star arrangement. However due to the condition $F_{d, N}^0 \times P_{d, N}^0 =(\rho_{\mathcal{V}})_{\mathbf{0}\mathbf{0}}$, the success probability in the star network is higher than the linear one (as shown in Fig. \ref{fig:number_vs_dimention_star_linear}), which, of course, follows from the previous observations.

\textbf{Note $\mathbf{1}$.} Unlike the case of a single resonator, attempting to cool multiple resonators with another auxiliary oscillator system does not show any prominent dimensional advantage in any configuration. In fact, the number of cycles of evolution and measurement required is still equal to the saturation values reported in Fig.~\ref{fig:number_vs_dimention_star_linear} giving rise to the same fidelity and success probability as in the case for $d = 6$.

\section{Cooling a hybrid system}
\label{sec:hybrid-cooling}

Going beyond the cooling of CV resonator systems, we will now exhibit that the measurement-based scheme can also be used to cool a CV-DV hybrid system comprising a single resonator, $\rho_\mathcal{V}$, and a \(d_s\)-level atom, $\rho_s$, which interact through a JC-type interaction. We aim to cool this hybrid system using another \(d\)-level atomic system, $\rho_R$, that acts as an auxiliary regulator. The regulator interacts directly with the oscillator, enabling energy exchange and dissipation. A hybrid system comprising an oscillator and a qubit, both in their ground state, $\ket{0}_\mathcal{V} \otimes \ket{0}_s$, can be used as the initial state to prepare several exotic states, as we shall delineate later. One of the immediate consequences of successfully cooling a CV-DV system is that we can employ the cooled resonator itself as a regulator for cooling of oscillators with a single round of evolution and conditional measurement as described in Sec.~\ref{sec:single_oscilator_cooling}. We can also use the cooled qudit as a regulator for the same purpose. {This means that starting with a single clean qudit in the state $\ket{k}_R$ (depending upon the desired measurement outcome) it is possible to cool entire networks of resonators by preparing other auxiliary regulator states.}

The Hamiltonian describing the dynamics of the CV-DV hybrid system, together with the auxiliary regulator, is given by
\begin{eqnarray}
\nonumber \hat{H} &=& \hat{a}_\mathcal{V}^\dagger \hat{a}_\mathcal{V} + \sum_{k=0}^{d_s-1} k \ketbra{k}{k}_s +\sum_{k=0}^{d-1} k \ketbra{k}{k}_R \\
\nonumber &+ &\lambda \hat{a}\Big( \sum_{k=0}^{d-1} \ketbra{k+1}{k} + \sum_{k=0}^{d_s-1} \ketbra{k_s+1}{k_s}\Big) \\
&+& \lambda \hat{a}^\dagger \Big(  \sum_{k=0}^{d-1} \ketbra{k}{k+1}+ \sum_{k=0}^{d_s-1} \ketbra{k_s}{k_s+1}\Big).
\label{equ:hybrid_gen_hamiltonian}
\end{eqnarray}
The first term in Eq.~\eqref{equ:hybrid_gen_hamiltonian} represents the free Hamiltonian of the CV resonator, $\rho_\mathcal{V}$, comprising the hybrid system, while the second term corresponds to the same for the constituent qudit, $\rho_s$, with internal energy levels, $\ket{k}_s$, and the third represents the auxiliary regulator, $\rho_R$, whose energy eigenstates are denoted by $\ket{k}_R$. The final two terms encompass the interaction between the oscillator and the two qudits. 

It is straightforward to observe that the aforementioned Hamiltonian too commutes with the total excitation operator, $\hat{\mathcal{N}}_e = \hat{a}_\mathcal{V}^\dagger \hat{a}_\mathcal{V} + \sum_{k=0}^{d_s-1} k \ketbra{k}{k}_s + \sum_{k=0}^{d-1} k \ketbra{k}{k}_R$, and thus preserves the total excitation number during the evolution. {As a result, the system dynamics are confined to invariant excitation-number subspaces, $\langle \hat{\mathcal{N}}_e \rangle = 0, 1, 2, \dots$, spanned by $\ket{n_\mathcal{V}, k_s, k_R}$ with $\langle \hat{\mathcal{N}}_e \rangle = n_\mathcal{V} + k_s + k_R$,} and similar to the previous sections, the Hamiltonian \(\hat{H}\), the time-evolution operator, \(\hat{U}(t)=e^{-\iota \hat{H} t}\), and the effective evolution operator for the hybrid system, $\hat{V}_{d}^k =~_d\bra{k} \hat{U}(t) \ket{k}_d$, acquire a block-diagonal structure. For our subsequent analysis, we focus primarily on measuring the regulator in its ground or excited-state, $\ket{0}_R, \ket{1}_R$, in each round of the protocol. Determining the optimal evolution time from the structure of $\hat{V}_d^{k \geq 2}$ while measuring on higher excited-states, $\ket{2 \leq k \leq d - 1}$, becomes analytically intractable.

In order to determine $t_{\text{opt}}^{k = 0,1}$, we need only to study the effective evolution operator in the $\langle \hat{\mathcal{N}}_e \rangle = 0, 1$ subspaces spanned by $\{\ket{0_\mathcal{V}, 0_s, 0_R}\}$ and $\{\ket{1_\mathcal{V}, 0_s ,0_R},\,\ket{0_\mathcal{V}, 1_s, 0_R},\,\ket{0_\mathcal{V}, 0_s, 1_R}\}$, respectively. The respective effective operators corresponding to repeated measurements in $\ketbra{0}{0}_R$ and $\ketbra{1}{1}_R$ respectively read

\begin{eqnarray}
\hat{V}^{0}_{d}
&=& \ketbra{0_\mathcal{V}, 0_s}{0_\mathcal{V},0_s} + e^{-\iota t}\Big[
\cos(\sqrt{2} t)\ket{1_\mathcal{V},0_s}\bra{1_\mathcal{V},0_s}
\nonumber\\
&&\quad
-\frac{\iota \sin(\sqrt{2} t)}{\sqrt{2}}
\big(\ketbra{1_\mathcal{V},0_s}{0_\mathcal{V},1_s}+\ketbra{0_\mathcal{V},1_s}{1_\mathcal{V},0_s}\big)
\nonumber\\
&&\quad
+\cos^{2}(\frac{ t}{\sqrt{2}})
\ket{0_\mathcal{V},1_s}\bra{0_\mathcal{V},1_s}
\Big] + \dots, ~~ \text{and} ~~ \label{eq:V0_hybrid}\\
\hat{V}^{1}_{d} &=& e^{-\iota t} \cos^{2}(\frac{ t}{\sqrt{2}}) \ketbra{0_\mathcal{V},0_s}{0_\mathcal{V},0_s} + \dots \label{eq:V1_hybrid}
\end{eqnarray}
The optimal evolution time corresponds to $t = t_{\text{opt}}^{k = 0, 1}$, at which the coefficient of $\ketbra{0_\mathcal{V}, 0_s}{0_\mathcal{V}, 0_s}$ in $\hat{V}_d^{k = 0, 1}$ becomes unity. It is evident from Eq.~\eqref{eq:V1_hybrid} that $t_{\text{opt}}^1 = \sqrt{2} \pi$ for measurement in the excited-state, $\ket{1}_R$. While the required coefficient is already unity in $\hat{V}_d^{0}$, the optimal time can be argued as the instance when the other remaining terms vanish, thereby reducing spurious contributions to the final fidelity and probability. We choose $t_{\text{opt}}^0 = \pi/\sqrt{2}$ whereby the $\sin(.)$ terms reduce to zero. Evolution through the optimal time intervals maximizes the survival probability of the projected state and optimizes the cooling performance.

\subsubsection{Dimensional advantage from qudit regulators in cooling the hybrid system}
\label{subsubsec:hybrid_results}
For numerical analysis of hybrid cooling, we consider the resonator to be in a displaced squeezed thermal state, $\rho_\mathcal{V} = \rho(\alpha = 0.5, r = 0.2, \bar{n} = 0.5)$, and the $d_s$ level qudit in a mixed state $\rho_s = \frac{1}{2}(\ketbra{0_s} +\mathbb{I}/{d_s})$, where \(\mathbb{I}\) denotes the $d_s$-dimensional identity matrix. The state $\rho_s$ naturally arises from imperfect state preparation, strong environmental decoherence (modeled by the depolarising channel, $\rho \to p \rho + (1 - p) \Tr(\rho) \mathbb{I}/d_s$ with $p = 1/2$), or unrecorded measurements acting on the atomic degrees of freedom, when trying to prepare it in the state $\ket{0}_s$~\cite{Lidar_arXiv_2019}.

\begin{figure}[ht]
\includegraphics[width=\linewidth]{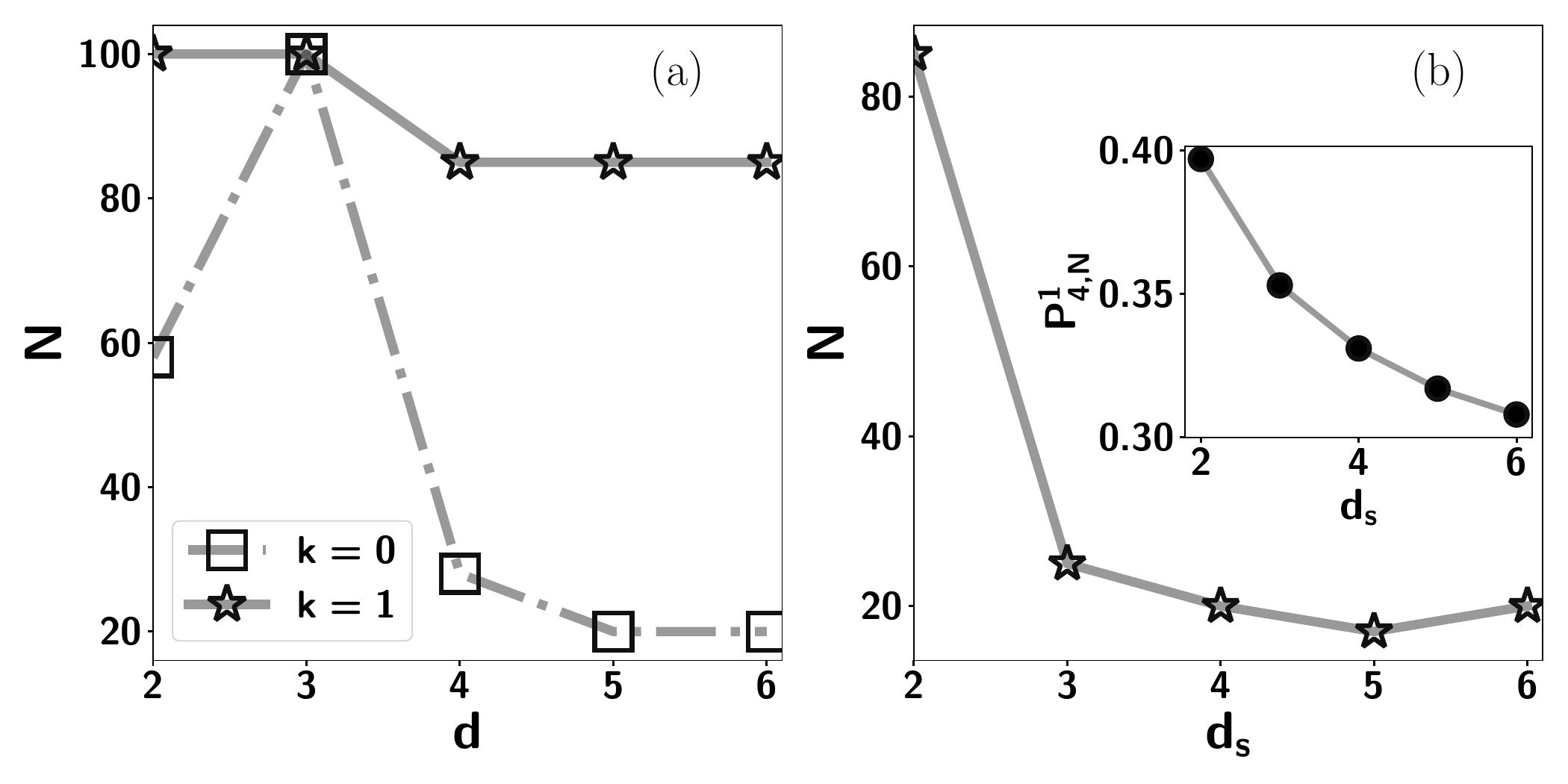}
\captionsetup{justification=Justified,singlelinecheck=false}
\caption{
{\bf Hybrid cooling of a displaced squeezed thermal resonator and qudit system coupled to a finite-dimensional auxiliary regulator:}
$(a)$ Number of evolution–measurement rounds, $N$ (ordinate), required to reach the saturation fidelity, against the regulator dimension, $d$ (abscissa), for post-selection on the ground, $k=0$ (hollow squares), and excited, $k=1$ (stars), states. Excited-state measurement requires more rounds but yields significantly higher final fidelity. 
$(b)$ Dependence of $N$ (ordinate) on the system dimension $d_s$ (abscissa) for fixed regulator dimension, $d=4$ under excited-state post-selection (\(\ket{k=1}_R\)). The
inset corresponds to the success probability $P^{1}_{4,N}$ (ordinate) with $d_s$ (abscissa). Increasing the auxiliary dimension reduces the required rounds while maintaining near-unit fidelity, demonstrating a dimensional advantage. All the axes are dimensionless. 
}
\label{fig:hybrid_aux_var} 
\end{figure}

{\it Role of measurement basis and dimension of regulator.}
Let us first consider $d_s=2$, i.e., the hybrid system comprises a harmonic oscillator and a qubit. 
\begin{table}[t]
\centering
\begin{tabular}{|c|c|c|c|c|c|c|}
\hline
$d$ & \multicolumn{2}{c|}{$N$} & \multicolumn{2}{c|}{$F_{d,N}^k$} & \multicolumn{2}{c|}{$P_{d, N}^k$} \\
\hline
    & $k = 0$    & $k = 1$    & $k = 0$        & $k = 1$        & $k = 0$         & $k = 1$        \\
\hline
$2$ & $58$       & $100$      & $0.664$        & $0.933$        & $0.595$         & $0.425$        \\
$3$ & $100$      & $100$      & $0.658$        & $0.950$        & $0.602$         & $0.418$        \\
$4$ & $28$       & $85$       & $0.665$        & $0.999$        & $0.596$         & $0.357$        \\
$5$ & $20$       & $85$       & $0.665$        & $0.999$        & $0.596$         & $0.357$        \\
$6$ & $20$       & $85$       & $0.665$        & $0.999$        & $0.596$         & $0.357$     \\
\hline
\end{tabular}
\caption{\justifying Fidelity, $F_{d, N}^k$, and probability, $P_{d, N}^k$, of successfully cooling the hybrid CV-DV system with $N$ rounds of evolution and measurement, conditioned on the level $\ketbra{k}{k}_R$ $(k = 0, 1)$, of the auxiliary qudit regulator of dimension, $d$. The initial states of the resonator, $\rho_\mathcal{V}$, and the qudit, $\rho_s$, constituting the hybrid system are specified in the main text.}
\label{tab:hybrid_d}
\end{table}
When the measurement outcome is $\ketbra{1}{1}_R$, the fidelity increases rapidly with the auxiliary dimension, $d$, and approaches unity for $d \ge 4$, indicating highly efficient purification of the target state. The required number of measurements decreases and saturates at $N = 85$, while the success probability is reduced. In contrast, for measurement in $\ketbra{0}{0}_R$, the fidelity saturates to a value less than unity to $0.66$ (see Table~\ref{tab:hybrid_d}). Interestingly, $N$ decreases with $d$ to achieve $F_{d, N}^0 \approx 0.66$ with a non-vanishing probability $P_{d, N}^0 \approx 0.6$ except $d=3$ as shown in Fig. \ref{fig:hybrid_aux_var}(a). There is still a dimensional advantage while going from $d = 2 \to 3$ in terms of a higher success probability, at the cost of a lower fidelity and an increased number of rounds. Our results show that the effectiveness of hybrid cooling strongly depends on the measurement basis of the auxiliary system, with {\it excited-state post-selection being significantly more effective than ground-state measurement.} When the auxiliary qudit is repeatedly measured in its ground state, the cooling protocol limits the purity of the prepared state. In contrast, when the auxiliary system is measured in its excited-state, the cooling fidelity increases with the dimension of the regulator and eventually saturates to unity. As a consequence, the excited-state measurement provides a near-perfect cooling of the hybrid system after a finite number of measurement steps.

{\it Dimensional benefit of the DV system in hybrid cooling.} It is also interesting to investigate whether increasing the dimension, $d_s$, of the DV system comprising the hybrid state also provides any dimensional advantage in the cooling process. Let us fix $d=4$ and analyze the case for excited-state measurement, since it has demonstrated superiority for $d_s = 2$. When the  resulting fidelity $F_{4, N}^1$ reaches close to unity, we notice \(N\) decreases drastically as \(d_s\) increases (see Fig. \ref{fig:hybrid_aux_var}(b)). This positive impact of the dimension of the cooled system can be obtained with a finite probability. It indicates that {\it near-perfect cooling is not only robust against changes in the system dimension, but also constructively affected by increasing $d_s$}: $N$ decreases with $d_s$ to yield $F_{4, N}^1 \to 1$ within an accuracy of $\mathcal{O}(10^{-3})$.

\subsection{Applications of the cooled hybrid system}
\label{subsec:hybrid_application}


If the cooling protocol successfully prepares the hybrid system near its ground-state manifold, the resulting CV-DV state can serve as a resource for generating exotic non-Gaussian CV states and hybrid entanglement between the two subsystems~\cite{Andersen_NP_2015_hybrid-system}. The cooling process initializes the system in a low-entropy, well-controlled state, providing a suitable starting point for subsequent operations. In this section, we will illustrate two applications of the hybrid state, $\ket{0_{\mathcal{V}}, 0_s}$ - $(1)$ synthesizing rotationally-symmetric Bosonic error correcting codes, and $(2)$ preparing hybrid maximally entangled as well as $N00N$ states.

\subsubsection{Creation of rotationally-symmetric Bosonic codes}
\label{subsubsec:error-codes}

While CV systems provide a multitude of advantages in quantum computation, e.g., by allowing for robust preparation of large-scale cluster states for measurement-based quantum computing~\cite{Nielsen_PRL_2004_optical-computing-cluster, Menicucci_PRL_2006_universal-CV-computation-cluster, Menicucci_PRL_2008_one-way-computing-frequency-comb, Flammia_JPB_2009_one-way-computer-frequency-comb, Gu_PRA_2009_computing-CV-cluster, Ohlinger_PRA_2012_measurement-based-CV, Menicucci_PRL_2014_fault-tolerant-CV-MBQC, Alexander_PRA_2016_one-way-computer-time-frequencyy-CV-cluster} and longer information-storage time~\cite{Ofek_Nature_2016_lifetime_qubit_superconducting, Campagne-ibarcq_Nature_2020_encoded-qubit_error-correction, Huang_Science_2022_quantum-advantage-experiment-long-coherence}, error correction to ensure fault-tolerance, inherently requires discretizing the Hilbert space. As such, CV error correcting codes are used to encode DV systems in the continuous Hilbert space, allowing for efficient error correction. One of the fundamental classes of CV error correcting codes comprises rotationally-symmetric codewords which admit a discrete rotation symmetry in the phase space. One such well-known and widely implemented error correcting code is the CAT code, which is a subclass of number-phase codes that allow for universal quantum computation using CV architectures~\cite {Grismo_PRX_2020_rotationa-symmetric-bosonic-codes}. Let us demonstrate how conditional CV operations, conditioned on the state of the qudit in the hybrid cooled state, can be used to generate these essential ingredients for fault-tolerant quantum computation.

The generic form of an even CAT state of $\mathcal{N}$ components, up to normalization, is given as

\begin{eqnarray}
    \ket{\Psi^\mathcal{N}}_{\text{CAT}} \sim \sum_{k = 0}^{\mathcal{N} - 1}\ket{\alpha \omega_{\mathcal{N}}^{k}} ~~ \text{with} ~~ \omega = e^{2 \pi \iota/\mathcal{N}}, \label{eq:even_CAT-N}
\end{eqnarray}
which is $+1$ eigenstate of the parity operator, $\hat{P} = e^{\iota \pi \hat{a}^\dagger \hat{a}}$, when $\mathcal{N}$ is even. We can prepare even CAT codewords for $\mathcal{N} = 2, 4, 6, \dots$ using the hybrid cooled system comprising qudits of even dimension. The first step is to apply the qudit Hadamard gate, defined as $\hat{H}_d \ket{j} = 1/\sqrt{d} \sum_{k = 0}^{d - 1} \omega_d^{jk} \ket{k}$, on $\ket{0_s}$ to generate the equal superposition $\ket{\Xi^d_s} = 1/\sqrt{d} \sum_{k = 0}^{d - 1} \ket{k_s}$ in the qudit subsystem. Defining the conditional displacement operator $\hat{D}_C = \sum_{k = 0}^{d - 1} \ketbra{k_s}{k_s} \otimes \hat{D}(\alpha \omega_\mathcal{N}^k)$, where the qudit is treated as the control, the $\mathcal{N} = d$ component even CAT codeword can then be prepared as

\begin{eqnarray}
    \ket{\Psi^\mathcal{N}}_{\text{CAT}} \sim \bra{\Xi^d_s}\hat{D}_C \Big( \frac{1}{\sqrt{d}} \sum_{k = 0}^{d - 1} \ket{k_s} \Big) \otimes \ket{0_\mathcal{V}} \ket{\Xi^d_s} ,\label{eq:CAT_preparation}
\end{eqnarray}
where the conditional displacement is followed by measuring the qudit in the Hadamard-rotated basis. The simplest example is for $\mathcal{N} = 2$, whence Eq.~\eqref{eq:CAT_preparation} becomes $\ket{\Psi^2}_{\text{CAT}} \sim \bra{+_s} (\ket{\alpha_\mathcal{V}} \otimes \ket{0_s} + \ket{-\alpha_{\mathcal{V}}} \otimes \ket{1_s}) \ket{+_s} \sim \ket{\alpha_\mathcal{V}} + \ket{-\alpha}_{\mathcal{V}}$, where we denote $\ket{+_s} = 1/\sqrt{2}(\ket{0_s} + \ket{1_s})$ for the qubit auxiliary regulator. Odd CAT states, i.e., the orthogonal logical codeword, which have $-1$ parity, can be generated from the even ones as $\hat{a}^{\dagger} \ket{\Psi^\mathcal{N}}_{\text{CAT}}$~\cite{Dakna_PRA_1997_odd-CAT_from_even, Ourjoumtsev_Science_2006_CAT-states}. 

\subsubsection{Generating hybrid entanglement and $N00N$ states}
\label{subsubsec:hybrid_ent}


Hybrid entangled (HE) states~\cite{Costanzo_PS_2015_hybrid-enatanglement_properties, Bose_PRAppl_2024_hybrid_entanglement-sharing} allow for the conversion of information between quantum processors operating over different encoding schemes. They can be used to increase the efficiency of Bell measurements in quantum teleportation in linear optical setups~\cite{He_SR_2022_hybrid_teleportation}, and have also been used as resources in remote state preparation~\cite{Morin_NP_2014_hybrid_remote-state-preparation} and quantum steering~\cite{Cavailles_PRL_2018_hybrid_steering, gefen_steering_1}. They allow for efficient distillation of CV Gaussian entanglement~\cite{Ourjoumtsev_PRL_2007_Gaussian_entanglement-distill, Zavatta_NP_2011_noiseless-amplifier, Datta_PRL_2012_compact_CV-distill} and reliable experimental demonstration of Bell inequality violation~\cite{Ji_PRL_2010_loophole-free_Bell}. On the computation front, they can be used for preparing the non-Gaussian cubic-phase and SNAP gates for universal CV quantum computation~\cite{Eriksson_NC_2024_cubic-phase-gate_experiment, Eickbusch_NP_2022_SNAP-gate}. Therefore, the generation of HE states is of paramount importance, and we describe how the cooled hybrid state, $\ket{0_\mathcal{V}, 0_s}$, can be employed for the same.

To prepare maximally entangled hybrid states of the form $\ket{\Psi}_{\text{HE}} \sim \sum_{k = 0}^{d - 1} \ket{k_\mathcal{V}, k_s}$, we may resort to a conditional photon addition scheme as
\begin{eqnarray}
   \ket{\Psi}_{\text{HE}} \sim \Big(\sum_{k = 0}^{d - 1} \ketbra{k_s}{k_s} \otimes \hat{a}_\mathcal{V}^{\dagger k} \Big) (\hat{H}_d \otimes \mathbb{I}_\mathcal{V}) \ket{0_s, 0_\mathcal{V}}, \label{eq:Hybrid_entanglement}
\end{eqnarray}
depending upon the dimension of the qudit subsystem. Note that heralded photon addition is an inherently probabilistic process, with the success probability reducing drastically with the number of photon-additions required. To improve the success probability, one can apply local squeezing operations on the CV mode before and after the photon addition protocol. Using $\hat{S}^\dagger(r) \hat{a}^{\dagger k}_\mathcal{V} \hat{S}(r) \ket{0}_\mathcal{V} = (\hat{a}_\mathcal{V}^\dagger \cosh r - \hat{a}_\mathcal{V} \sinh r)^k \ket{0}_\mathcal{V} = \cosh^k r \sqrt{(k+1)!} \ket{k}_\mathcal{V}$, we can prepare non-maximally HE states, $\ket{\Psi}_{\text{HE}} \sim \sum_{k = 0}^{d - 1} \cosh^k r \ket{k_\mathcal{V}, k_s}$,  in an attempt to improve the overall success probability and then perform entanglement distillation to obtain highly entangled states.

On the other hand, $N00N$ states~\cite{Dowling_CP_2008_NOON-state} find vast applications in both single- and multi-parameter quantum metrology~\cite{Hosler_PRA_2013_NOON-state_estimation, Slussarenko_NP_2017_NOON-state_estimation, You_APR_2021_NOON-state_estimation, Hong_NC_2021_NOON-state_multi-estimation, Namkung_NJP_2024_NOON-state_multi-estimation}, as well as in linear-optical quantum computation~\cite{Franson_PRL_2002_NOON-state_LOQC, Ralph_PRA_2002_NOON-state_CNOT, Kok_RMP_2007_LOQC_review} and quantum memories~\cite{Zhang_PRA_2018_NOON-state_memory}. When the qudit subsystem comprising the hybrid cooled state is of dimension $d$, we can prepare $N00N$ states of the form $\ket{\Psi}_{N00N} \sim \ket{N = (d - 1)_s, 0_\mathcal{V}} + \ket{0_s, N = (d - 1)_\mathcal{V}}$ by first using a subspace rotation operator to convert the qudit as $\ket{0}_s \to 1/\sqrt{2}(\ket{0}_s + \ket{d - 1}_s)$ and then applying the conditional photon-addition with the qudit as the control

\begin{eqnarray}
    \nonumber \ket{\Psi}_{N00N} &\sim& \Big( \ketbra{0_s}{0_s} \otimes \hat{a}_\mathcal{V}^{\dagger (d - 1)} \\
    \nonumber &+& \ketbra{(d - 1)_s}{(d - 1)_s} \otimes \mathbb{I}_\mathcal{V} \Big) \\
    &\frac{1}{\sqrt{2}}&(\ket{0}_s + \ket{d - 1}_s) \otimes \ket{0}_\mathcal{V}. \label{eq:N00N}
\end{eqnarray}
Once again, we can use heralded photon-addition interleaved with local squeezing to obtain a better success probability at the expense of preparing an inexact $N00N$ state.

While the aforementioned protocols for error-correcting codeword generation and HE state preparation are well known and implemented, they still require the initial creation of clean hybrid systems in their ground state. Our cooling protocol thus provides the initial necessary ingredient for such processes - we can start with CV and DV systems in mixed states of low purity and convert them to perfect vacuua before beginning the resource-generation schemes.

\section{Conclusion}
\label{sec:conclu}

Hybrid quantum systems comprising both discrete variable (DV) and continuous variable (CV) degrees of freedom allow for the exchange of information between different platforms, offering significant advantages in a broad range of quantum information–processing tasks, spanning from quantum key distribution~\cite{Djordjevic_IEEE_2020_hybrid_key-distribution} to the near-deterministic implementation of quantum gates~\cite{Nemoto_PRL_2004_hybrid_CNOT, Spiller_NJP_2006_hybrid_computation}. They form the cornerstone of superconducting setups, which are attracting increasing attention in the field of universal~\cite{Gottesman_PRA_2001_GKP-code, Heeres_PRL_2015_SNAP, Krastanov_PRA_2015_SNAP, Hillmann_PRL_2020_cubic-phase-gate, Eriksson_NC_2024_cubic-phase-gate_experiment, Kundra_PRX_2022_SNAP-displacement_state-preparation} fault-tolerant~\cite{Barends_Nature_2014_surface-code_superconducting, Ofek_Nature_2016_lifetime_qubit_superconducting, Rosenblum_Science_2018_fault-tolerance_superconducting, Hu_NP_2019_binomial-code_superconducting, Cai_FR_2021_error-correction_superconducting} quantum computation. While initially proposed with qubits constituting the DV subsystem, the dimensional advantage offered by higher-dimensional discrete systems or qudits motivates the study of hybrid oscillator-qudit systems.


Starting from an ensemble of oscillators prepared in the most general single-mode mixed Gaussian state, namely a displaced squeezed thermal state, we investigated a refrigeration protocol that cools these states to vacuum by employing an auxiliary qudit, repeated unitary evolution, and subsequent projective measurements. To establish the necessity of higher-dimensional regulators, we proved a no-go theorem, showing that
a two-level auxiliary regulator system cannot cool a CV resonator with near-unit fidelity and a non-vanishing success probability within this framework. For a single resonator, we derived the optimal interaction times corresponding to different regulator dimensions and showed that they are independent of the initial state of the target system. Further, we identified a two-fold dimensional advantage of qudit auxiliaries:  increasing the regulator dimension significantly reduces the number of evolution–measurement cycles required for convergence, and enables oscillators with higher initial energies to be cooled more efficiently.  In addition, we extended our analysis to the simultaneous cooling of multiple resonators arranged in two distinct configurations, viz, linear and star networks. While near-perfect cooling with a finite success probability is achievable for linear networks, the protocol fails for star networks, even for small systems consisting of only two or three oscillators. Moreover, we showed that although increasing the regulator dimension initially enhances cooling performance, unbounded dimensional growth offers no further benefit; the dynamics saturate to state-dependent limits for moderate regulator dimensions, indicating an intrinsic bound on the effectiveness of the cooling mechanism.

We furthermore examined the cooling dynamics of a hybrid CV–DV system mediated by an auxiliary qudit. Our results demonstrate near-unit cooling fidelity with a moderate success probability for a regulator of dimension $d \geq 4$. We further identified a collateral benefit from increasing the discrete target subsystem dimension in the hybrid architecture, which leads to a reduced number of interaction–measurement cycles required to achieve effective cooling. Beyond refrigeration, we showed that the proposed protocol can be exploited for the generation of non-Gaussian resources, including Schr{\"o}dinger CAT states and hybrid entangled states such as the $N00N$ state. Since the direct preparation of pure low-entropy states remains experimentally challenging, yet essential for a plethora of information-theoretic protocols, our cooling mechanism may offer a practically appealing route by providing a natural and versatile physical platform for such state engineering tasks.

\acknowledgements

We acknowledge discussions with Tanoy Kanti Konar. We acknowledge the cluster computing facility at Harish-Chandra Research Institute and the use of QIClib -- a modern C++ library for general-purpose quantum information processing and quantum computing~\cite{QIClib}. 
We acknowledge support from the project entitled ``Technology Vertical - Quantum Communication'' under the National Quantum Mission of the Department of Science and Technology (DST)  (Sanction Order No. DST/QTC/NQM/QComm/$2024/2$ (G)). R. G. acknowledges funding from the HORIZON-EIC-$2022$-PATHFINDERCHALLENGES-$01$ program under Grant Agreement No.~$10111489$ (Veriqub). Views and opinions expressed are those of the authors only and do not necessarily reflect those of the European Union. Neither the European Union nor the granting authority can be held responsible for them.

\appendix

\section{Continuous variable Gaussian systems - the bare minimum}
\label{sec:app1}

Continuous variable (CV) systems live in an infinite-dimensional Hilbert space and are characterized by the position, $\hat{x}$, and momentum, $\hat{p}$, quadrature operators, which obey the Bosonic canonical commutation relation (CCR), $[\hat{x}, \hat{p}] = \iota$ (we assume $\hbar = 1$)~\cite{Adesso_OSID_2014, Serafini_2017}. The annihilation, $\hat{a}$, and creation, $\hat{a}^\dagger$, operators are further defined as $\hat{a}(\hat{a}^\dagger) = \frac{\hat{x} + (-) \iota \hat{p}}{\sqrt{2}}$ obeying $[\hat{a}, \hat{a}^\dagger] = 1$. The free Hamiltonian for any CV system is that of the simple harmonic oscillator, $\hat{H}_{\text{free}} = (\hat{x}^2 + \hat{p}^2)/2 = (\hat{a}^\dagger \hat{a} + 1/2)$.  For multipartite CV systems, also known as multimode CV systems, the corresponding quadrature operators can be grouped into the vector $\hat{\mathbf{R}} = (\hat{x}_1, \hat{p}_1, \dots, \hat{x}_m, \hat{p}_m)$ where $m$ represents the total number of constituent modes. Defining $\Omega_m = \oplus_{i = 1}^m \begin{pmatrix}
    0 & 1 \\
    -1 & 0
\end{pmatrix}_i$ as the $m$-mode symplectic form, the CCR can be succinctly represented as $[\hat{\mathbf{R}}, \hat{\mathbf{R}}^T] = \iota \Omega_m$. The simplest CV quantum states are Gaussian states, defined as the ground and thermal states of Hamiltonians of the form $\hat{H} = \hat{\mathbf{R}}^T H \hat{\mathbf{R}} + \hat{\mathbf{R}}^T \mathbf{r}$, that are at most second order in the quadrature operators~\cite{ferraro2005, Weedbrook_RMP_2012}. Here $\mathbf{r}$ is a $2 \times m$-dimensional real vector (for $m$ modes) and $H > 0$ is a positive-definite symmetric matrix. As the name suggests, Gaussian states have Gaussian wavefunctions, and can be characterized solely by the first and second moments of the vector $\hat{\mathbf{R}}$, known respectively as the displacement vector and the covariance matrix, defined as

\begin{eqnarray}
    \mathbf{d} &=& \langle \hat{\mathbf{R}} \rangle = (\langle \hat{x}_1 \rangle, \langle \hat{p}_1 \rangle, \dots, \langle \hat{x}_m \rangle, \langle \hat{x}_m \rangle)^T \label{eq:app1_displacement} \\
    \sigma &=& \langle \{(\hat{\mathbf{R}} - \mathbf{d}),(\hat{\mathbf{R}} - \mathbf{d})^T\} \rangle. \label{eq:app1_covariance}
\end{eqnarray}
Here, $\{\hat{A}, \hat{B}\} = \hat{A}\hat{B} + \hat{B}\hat{A}$ denotes the anticommutator of the two operators, $\hat{A}$ and $\hat{B}$. Therefore, even though they are essentially infinite-dimensional systems, any $m$-mode Gaussian state can be represented by a $2m$-dimensional displacement vector, $\mathbf{d}$, and a $2m \times 2m$-dimensional real symmetric covariance matrix, $\sigma$. In this picture, any $m$-mode quadratic Hamiltonian can be written as $\hat{\mathcal{H}}=\frac{1}{2} \hat{\xi}^\dagger \hat{H} \hat{\xi}$ with $\hat{\xi}=\left(\hat{a}_1, \hat{a}_2,..., \hat{a}_m, \hat{a}_1^\dagger,..., \hat{a}_m^\dagger\right)^T$. We can construct the symplectic matrix, $S_H$, corresponding to $\hat{H}$ as~\cite{Luis_QSO_1995, Arvind_Pramana_1995, Adesso_OSID_2014}
 \begin{eqnarray}
     S_H = T^\dagger L^\dagger \exp{- i K \hat{H} } L T,
     \label{eq:H_symp}
 \end{eqnarray}
 where $K, L$ and $T$ are $2m \times 2m$ matrices given by 
 \begin{eqnarray}
    && K = \begin{pmatrix}
         \mathbb{I}_m & \mathbb{O}_m \\
         \mathbb{O}_m & -\mathbb{I}_m
     \end{pmatrix}, \label{eq:K_mat} \\
    && L = \frac{1}{\sqrt{2}}\begin{pmatrix}
         \mathbb{I}_m & \iota \mathbb{I}_m \\
         \mathbb{I}_m & -\iota \mathbb{I}_m
     \end{pmatrix}, \label{eq:L_mat} \\
    && T_{jk} = \delta_{k,2j - 1} + \delta_{k + 2m, 2j} \label{eq:T_mat},
 \end{eqnarray}
with $\mathbb{I}_m$ being the $m$-dimensional identity and $\mathbb{O}_m$ being the null matrix.
The evolved Gaussian state, in terms of its displacement vector and covariance matrix, can be characterized as \cite{Adesso_OSID_2014}
\begin{eqnarray}
&& \nonumber \mathbf{d} \to \mathbf{d}' = S_H \mathbf{d}, \label{eq:app1_evolved_disp} ~\text{and}~ \\
&& \sigma \to \sigma' = S_H \sigma S_H^T, \label{eq:app1_evolved_cov}.
\end{eqnarray}
It is important to note that the dynamics can be described in its entirety by the above formalism only if all components, viz, the initial state as well as the evolution Hamiltonian, are Gaussian.

\bibliography{ref}

\end{document}